\documentclass[10pt,aps,prd,twocolumn,floatfix,nofootinbib,superscriptaddress,
preprintnumbers]{revtex4-2}

\usepackage{amsmath,amssymb,bm}
\usepackage{graphicx}
\usepackage{color}
\definecolor{orange}{rgb}{1,0.5,0}
\usepackage{hyperref}
\hypersetup{
   hidelinks,
   pdftitle={Holographically motivated Hubble-running dark energy: Phantom-divide crossing in VCDM and observational constraints},
   pdfauthor={Antonio De Felice, Burin Gumjudpai, Shinji Tsujikawa, and Jack Zidich}
}
\usepackage{enumitem}
\usepackage{array}[=2016-10-06]
\usepackage{mathrsfs,booktabs}

\numberwithin{equation}{section}
\renewcommand{\theequation}{\arabic{section}.\arabic{equation}}
\newcommand{\mpl}{M_{\rm Pl}}
\newcommand{\Ode}{\Omega_{\rm DE}}
\newcommand{\Or}{\Omega_r}
\newcommand{\Om}{\Omega_m}
\newcommand{\chde}{c_{\rm hde}}
\newcommand{\dd}{\mathrm{d}}
\newcommand{\RaggedCell}[1]{{\raggedright #1\par}}
\newcommand{\N}{N}
\newcommand{\be}{\begin{equation}}
\newcommand{\ee}{\end{equation}}
\newcommand{\ba}{\begin{eqnarray}}
\newcommand{\ea}{\end{eqnarray}}

\begin{document}
\preprint{YITP-26-151, WUCG-26-11}

% Both preprint numbers are retained from the supplied grqcv6.tex.

\title{Holographically motivated Hubble-running dark energy:\\
Phantom-divide crossing in VCDM and observational constraints}

\author{Antonio De Felice}
\email{antonio.defelice@yukawa.kyoto-u.ac.jp}
\affiliation{Center for Gravitational Physics and Quantum Information,
Yukawa Institute for Theoretical Physics, Kyoto University, 606-8502, Kyoto, Japan}

\author{Burin Gumjudpai}
\email{burin.gum@mahidol.ac.th}
\affiliation{NAS, Centre for Theoretical Physics and Natural Philosophy, Mahidol University,
Nakhonsawan Campus, Nakhonsawan 60130, Thailand}

\author{Shinji Tsujikawa}
\email{tsujikawa@waseda.jp}
\affiliation{Department of Physics, Faculty of Science and Engineering,
Waseda University, Shinjuku, Tokyo 169-8555, Japan}

\author{Jack Zidich}
\email{j.zidich@yukawa.kyoto-u.ac.jp}
\affiliation{Center for Gravitational Physics and Quantum Information,
Yukawa Institute for Theoretical Physics, Kyoto University, 606-8502, Kyoto, Japan}

\date{\today}

\begin{abstract}
Motivated by power-law dependences on the expansion scale in holographic
dark-energy constructions, we study the Hubble-running equation of state
$w_{\rm DE}=w_0(H/H_0)^q$ (HoloEoS), where $H$ is the Hubble expansion rate,
$H_0$ and $w_0$ are the present values of $H$ and $w_{\rm DE}$, and $q$ is
a constant dimensionless exponent controlling the running.
We embed this history in VCDM, a type-II minimally modified gravity theory
that propagates only the two tensor modes of general relativity.
For regular expanding histories with standard matter, we reconstruct
the potential $V(\phi)$
of its nondynamical auxiliary scalar field, providing a Lagrangian
description of both the homogeneous background and cosmological
perturbations. This framework allows HoloEoS to realize a smooth late-time
crossing of the phantom divide $w_{\rm DE}=-1$.
We confront this VCDM realization with SPT, Planck, and ACT CMB data,
updated DESI DR2 distances, and optionally DESY5 supernovae with Dovekie
recalibration, comparing it with $\Lambda$CDM and the
Chevallier--Polarski--Linder parameterization embedded in the same
gravitational framework (CPL-VCDM).
For the full data combination, the posterior medians and central
$68\%$ credible intervals are
$w_0=-0.881^{+0.040}_{-0.038}$ and $q=0.453^{+0.143}_{-0.133}$,
with a crossing redshift $z_\times=0.476^{+0.106}_{-0.081}$
conditional on $0<z_\times<6$.
For this data combination, physical-likelihood optimization by
Metropolis--Hastings (MH) simulated annealing gives Akaike information
criterion differences of $\Delta\mathrm{AIC}\simeq-3.40$ for HoloEoS
and $-4.10$ for CPL-VCDM relative to $\Lambda$CDM.
Thus, a Hubble-scale-motivated dark-energy history realized within
a specified gravitational theory achieves fitting performance comparable
to CPL-VCDM while naturally allowing phantom-divide crossing.
\end{abstract}

\maketitle

%%%%%%%%%%%%
\section{Introduction}
\label{sec:introduction}
%%%%%%%%%%%%

The late-time accelerated expansion of the Universe was first established
through observations of Type~Ia supernovae~\cite{RiessDE1998,PerlmutterDE1999}.
Subsequent measurements of cosmic microwave background (CMB)
anisotropies~\cite{WMAP:2012nax,PlanckCosmology2018}
and baryon acoustic oscillations (BAO)~\cite{SDSS:2005xqv}
have provided complementary evidence.
The spatially flat $\Lambda$CDM model, which includes a cosmological
constant and cold dark matter (CDM), provides a simple and successful
description of these observations.
However, it does not explain why the observed vacuum-energy density
is so small compared with quantum-field-theory estimates or why
the matter and dark-energy (DE) densities are comparable
today~\cite{WeinbergCC1989,PeeblesRatra2003}.
These unresolved questions motivate studies of dynamical DE and
modified gravity; see
Refs.~\cite{CopelandDE2006,SilvestriTrodden:2009,CliftonEtAl:2012,
TsujikawaQuintessence2013,JoyceCosmology2015,Koyama:2016,
BullEtAl:2016,Heisenberg:2019Review,KaseTsujikawa:2019Review}
for representative reviews.

Recent BAO measurements from the Dark Energy Spectroscopic Instrument
(DESI)~\cite{DESI:2024mwx}, particularly those from Data Release~2
(DR2)~\cite{DESI:2025DR2BAO}, provide further motivation for exploring
dynamical DE.
A widely used description of the DE equation-of-state parameter is the
Chevallier--Polarski--Linder (CPL)
parametrization~\cite{ChevallierPolarski2001,LinderExpansion2003},
$w_{\rm DE}(a)\equiv P_{\rm DE}/\rho_{\rm DE}=w_0+w_a(1-a)$.
Here $P_{\rm DE}$ and $\rho_{\rm DE}$ are the DE pressure and energy
density, respectively; $a$ is the scale factor normalized to unity today,
with $1+z=a^{-1}$; $w_0$ is the present value of $w_{\rm DE}$; and $w_a$
is a constant parameter describing its evolution.
Within this parametrization, joint analyses of DESI DR2, CMB, and
Type~Ia supernova data favor the region $w_0>-1$ and $w_a<0$, with the
strength of the preference for evolving DE depending on the supernova
sample~\cite{DESI:2025DR2BAO}.
Representative best-fit histories cross the phantom divide
$w_{\rm DE}=-1$ at low redshift, evolving from $w_{\rm DE}<-1$ in the
past to $w_{\rm DE}>-1$ today.

Realizing such a crossing in a consistent theory is nontrivial.
In general relativity (GR), a minimally coupled canonical scalar field with
positive energy density cannot enter the phantom regime
$w_{\rm DE}<-1$, whereas reversing the sign of its kinetic term
introduces a ghost~\cite{CaldwellPhantom2002,CarrollPhantom2003,
Cline:2003gs}.
More generally, a regular crossing with stable perturbations is
obstructed in minimally coupled single-field
k-essence~\cite{VikmanCrossing2005}.

Consistent phantom-divide crossing can arise through several mechanisms
in extended DE and modified-gravity theories.
Nonminimal scalar--curvature couplings provide one route and have been
investigated in connection with the DESI results in
Refs.~\cite{Ye:2024desi,Wolf:2024nmc,Ye:2024tg,Pan:2025desi,
Wolf:2025nmc,Adam:2025nmc,SanchezLopez:2025nmc,ArgudoPanes:2026nmc}.
Such couplings generally induce time variation of the locally measured
effective gravitational coupling $G_{\rm eff}$.
This time variation can persist even when the Vainshtein mechanism
efficiently suppresses the scalar-mediated fifth force in local
regions~\cite{Babichev:2011iz,Kimura:2011dc}. 
Their cosmological evolution must therefore satisfy stringent local
gravity bounds, particularly those from lunar laser
ranging~\cite{Hofmann:2018myc,Tsujikawa:2019pih}.

Scalar derivative interactions provide a different route: a potential
that breaks shift symmetry can support stable crossing in a Galileon
model~\cite{Tsujikawa:2025wca}.
A related kinetically braided model with dark-sector momentum exchange
realizes stable crossing with suppressed structure
growth~\cite{PookkillathTsujikawa2026}.
Extending generalized-Proca theories~\cite{Heisenberg:2014rta,
Tasinato:2014eka,BeltranJimenez:2016rff} with a canonical scalar field
and its potential yields scalar--vector--tensor (SVT) models that
can realize phantom-divide crossing~\cite{Tsujikawa:2026xqm}. 
An analysis using Dark Energy Survey (DES) Year~5 supernovae,
DESI DR2 BAO, compressed CMB
information, and redshift-space-distortion measurements finds this
realization consistent with expansion and structure-growth data, with a
sampled best-fit improvement of $\Delta\chi^2\simeq-12$ relative to flat
$\Lambda$CDM~\cite{RoyEtAl:2026SVTConstraints}.

Most of the models discussed above rely on dynamical scalar or vector
fields to modify the DE evolution.
Constrained modifications of gravity offer an alternative by modifying
the constraint structure to eliminate extra propagating gravitational
degrees of freedom.
An early example is the cuscuton~\cite{AfshordiCuscuton2007,
AfshordiCuscutonCosmology2007}, whose scalar perturbation obeys a
constraint equation on cosmological backgrounds and does not describe
an independent propagating mode.
Another example is VCDM, a type-II minimally modified gravity theory
that propagates only two local tensor degrees of
freedom~\cite{VCDMFoundation2020}.
The potential $V(\phi)$ of its nondynamical auxiliary scalar $\phi$
determines the effective DE evolution and can accommodate
phantom-divide crossing~\cite{VCDMBorghetto2026}.

The holographic principle offers a complementary motivation for
constructing DE models. It suggests that the number of independent
degrees of freedom in a gravitating region is controlled by its
boundary area rather than its
volume~\cite{Hooft:1993gx,Susskind:1994vu,Bousso:2002ju}.
Cosmological holographic DE (HDE) rests more specifically on the
Cohen--Kaplan--Nelson ultraviolet/infrared (UV/IR) energy
bound~\cite{Cohen:1998zx}.
Consider an effective field theory in a region of linear size $L$,
which sets the infrared scale. With an ultraviolet energy cutoff
$\Lambda_{\rm UV}$, the total energy is of order $L^3\Lambda_{\rm UV}^4$.
The mass of a black hole of the same size is of order $L\mpl^2$.
Requiring the total energy not to exceed this mass gives
\begin{equation}
   L^3\Lambda_{\rm UV}^4\lesssim L\mpl^2.
   \label{eq:CKNbound}
\end{equation}
If this inequality is saturated and the vacuum contribution is identified
with DE, one obtains
\begin{equation}
   \rho_{\rm hde}=3\chde^2\mpl^2L^{-2},
   \label{eq:HDE_density}
\end{equation}
where $\mpl\equiv(8\pi G)^{-1/2}$ is the reduced Planck mass, $G$ is
Newton's gravitational constant, $\rho_{\rm hde}$ is the HDE energy density,
and $\chde>0$ is a time-independent dimensionless constant whose value is
not fixed by the bound.
Saturation of the bound and the choice of the infrared length are
additional model assumptions.

The cosmological evolution of HDE depends on the infrared cutoff $L$.
We denote the Hubble expansion rate by $H\equiv\dot a/a$, where a dot
denotes differentiation with respect to cosmic time $t$.
In the standard noninteracting model with pressureless matter, the
choice $L=H^{-1}$ gives $w_{\rm DE}=0$ and therefore cannot drive
late-time acceleration~\cite{Hsu:2004ri}.
Accelerating solutions can instead be obtained by choosing the future
event horizon as the infrared cutoff~\cite{Li:2004rb}, or by relating
$L^{-2}$ to a local combination of $H^2$ and
$\dot H$~\cite{Gao:2007ep,Granda:2008dk,Granda:2008tm}.
Another approach modifies the horizon-entropy relation, yielding HDE
densities that scale as nonstandard powers of
$L^{-1}$~\cite{Tavayef:2018xwx,Saridakis:2020zol}.
For a Hubble-radius cutoff, this gives a power-law dependence on $H$.
These constructions are summarized in Appendix~\ref{app:representativeHDE}.

Hubble-dependent and inhomogeneous DE equations of state, together with
generalized infrared-cutoff prescriptions, have been investigated
previously~\cite{NojiriInhomogeneous2005,Nojiri:2005pu,Nojiri:2017opc}.
Motivated by these power-law scalings, we consider the two-parameter
ansatz
$w_{\rm DE}=w_0(H/H_0)^q$, which we call HoloEoS.
Here $H_0$ and $w_0$ are the present values of $H$ and $w_{\rm DE}$,
respectively, and $q$ is a constant dimensionless exponent.
This phenomenological ansatz is inspired by holographic models but is
not uniquely derived from a particular infrared cutoff.
We realize HoloEoS within VCDM~\cite{VCDMFoundation2020} by reconstructing
the potential $V(\phi)$ for the corresponding expansion history.

Early cosmological applications of VCDM used this reconstruction freedom
to realize a rapid transition at low redshift~\cite{VCDMH0DeFelice2021}.
Subsequent studies constructed smooth realizations of sign-switching
cosmological-constant models~\cite{VCDMLambdaSAkarsu2026},
constrained standard dynamical-DE parametrizations, including CPL,
with observational data~\cite{VCDMArora2025}, and investigated
phantom-divide crossing with an extended neutrino
sector~\cite{VCDMNeutrinosLadeira2026}.
A data-driven analysis combining nonparametric reconstruction
with symbolic regression identified the one-parameter form
$w(a)=w_0/\sqrt{a}$~\cite{VCDMBorghetto2026}.
HoloEoS ties the equation-of-state evolution directly to the
normalized Hubble rate $H/H_0$.

Specifying the perturbation dynamics is essential because models with
the same homogeneous background geometry can predict different CMB
anisotropies, gravitational lensing, and matter growth.
If DE is modeled as a separately conserved, purely adiabatic fluid,
its pressure perturbation satisfies
$\delta P_{\rm DE}=c_{a,\rm DE}^2\delta\rho_{\rm DE}$, where
$c_{a,\rm DE}^2\equiv\dot P_{\rm DE}/\dot\rho_{\rm DE}$.
The coefficient $c_{a,\rm DE}^2$ is generically singular at
phantom-divide crossing, but its divergence does not by itself imply
singular physical perturbations~\cite{KunzSapone:2006}.
In our VCDM realization, effective DE arises from the gravitational
sector without an independent DE fluid.
The metric and standard-matter perturbations follow from the VCDM
field equations without imposing the above fluid prescription or
identifying $c_{a,\rm DE}^2$ with a physical propagation speed.
These equations remain regular through the crossing on a smooth
cosmological background~\cite{VCDMFoundation2020,VCDMArora2025}.
Our likelihood analysis therefore tests this specific gravitational
realization of HoloEoS.

Alongside the expansion history, we examine the inferred values of
$H_0$, the present matter density fraction $\Omega_{m0}$, and the clustering
combination $S_8\equiv\sigma_8\sqrt{\Omega_{m0}/0.3}$.
Here $\sigma_8$ is the root-mean-square amplitude of the linear matter
density contrast smoothed with a spherical top-hat window of
comoving radius $8h^{-1}\,{\rm Mpc}$, and
$h\equiv H_0/(100\,{\rm km}\,{\rm s}^{-1}\,{\rm Mpc}^{-1})$ is the
dimensionless Hubble constant.
Throughout this paper, $\sigma_8$ and $S_8$ denote their present-day
($z=0$) values.

We derive the HoloEoS phantom-divide crossing condition and show how the
reconstructed VCDM potential remains regular through the crossing.
We compare HoloEoS with $\Lambda$CDM and CPL realized in the same
VCDM framework (CPL-VCDM), adopting a common neutrino sector.
Using South Pole Telescope (SPT), Planck, and Atacama Cosmology
Telescope (ACT) CMB measurements, updated DESI DR2 distances,
and optionally DESY5 supernovae with Dovekie recalibration, we find
physical-likelihood improvements comparable to those of CPL-VCDM.
For the full data combination, MH simulated-annealing
optimization gives Akaike information criterion
(AIC)~\cite{Akaike:1974} differences relative to $\Lambda$CDM of
$\Delta\mathrm{AIC}\simeq-3.40$ and $-4.10$, respectively.

This paper is organized as follows.
Section~\ref{sec:II} presents the background equations and the
reconstruction of the VCDM potential.
Section~\ref{sec:holoeos} introduces HoloEoS and examines its background
evolution and phantom-divide crossing.
Section~\ref{sec:constraints} describes the data and methods, presents
observational constraints, and compares HoloEoS with $\Lambda$CDM
and CPL-VCDM.
Section~\ref{sec:conclusions} summarises our findings and their
physical implications.
Appendix~\ref{app:representativeHDE} reviews the holographic DE
constructions motivating HoloEoS.
Appendix~\ref{app:vcdm_perturbations} provides the VCDM perturbation
equations and initial conditions.
Appendix~\ref{app:numerical_checks} details the likelihood
conventions, numerical checks, and annealing procedure.

\section{General setup and VCDM dictionary}
\label{sec:II}
\subsection{Universal background equations}
\label{subsec:universalBackground}

The analytic equations in this subsection assume that the standard
sector consists of two separately conserved components: pressureless
matter and massless radiation.
The observational analysis in Sec.~\ref{sec:constraints} includes
massive thermal neutrinos, whose density and pressure are retained
in the physical CLASS background and in the sums over standard species.
The neutrino prescription and numerical implementation are specified
in Sec.~\ref{subsec:run_settings}.

We first describe the target background using the GR equations for
separately conserved effective DE, pressureless matter, and radiation.
The matter and radiation sectors are minimally coupled to the metric,
and there is no energy exchange between the three components.
This effective-fluid description is used only to specify the homogeneous
history. In Sec.~\ref{subsec:VCDMreconstruction}, the same history is
generated within VCDM by the potential $V(\phi)$, without introducing
an independent DE matter field.

We consider a spatially flat Friedmann--Lema\^itre--Robertson--Walker (FLRW)
spacetime with line element
\begin{equation}
 \dd s^2=-\dd t^2+a^2(t)\delta_{ij} 
 {\rm d} x^i {\rm d} x^j\,,
 \label{eq:flatFLRWmetric}
\end{equation}
where $t$ is cosmic time, $a(t)$ is the scale factor, 
and $x^i$ denotes comoving Cartesian coordinates.  
The Hubble expansion rate is
$H\equiv\dot a/a$, and an overdot denotes differentiation with respect to
$t$.  We use natural units, $c=\hbar=1$, unless otherwise specified.

We denote by $\rho_{\rm DE}$, $\rho_m$, and $\rho_r$ the energy densities
of the effective DE component, pressureless matter comprising baryons 
and CDM, and radiation, respectively.  In terms of the reduced
Planck mass $\mpl$, the corresponding fractional energy densities are
defined by
\be
 \Omega_{\rm DE} \equiv \frac{\rho_{\rm DE}}{3\mpl^2H^2}, 
 \qquad
 \Omega_{i} \equiv \frac{\rho_i}{3\mpl^2H^2}, 
\ee
where $i=m,r$.

The pressure associated with $\rho_{\rm DE}$ is denoted by
$P_{\rm DE}$, and the corresponding DE equation-of-state parameter is
defined by
\begin{equation}
 w_{\rm DE}\equiv\frac{P_{\rm DE}}{\rho_{\rm DE}}.
\end{equation}
We normalize the present scale factor to $a_0=1$ and introduce the
e-fold number $\N\equiv\ln a$ and the redshift $1+z=a^{-1}$.
A subscript $0$ denotes a present-day value.
In particular, present-day fractional density parameters are written as
$\Omega_{I0}\equiv\Omega_I(z=0)$ for a density component $I$,
for example $\Omega_{b0}$, $\Omega_{m0}$, and
$\Omega_{{\rm DE},0}$.
Density fractions without the subscript $0$ denote time-dependent
quantities.
On a spatially flat background, the Friedmann constraint reads
\begin{equation}
 \Om=1-\Ode-\Or.
 \label{eq:flatconstraint}
\end{equation}
For pressureless matter and radiation, we take $P_m=0$ and
$P_r=\rho_r/3$, respectively. Their separate conservation equations then
yield $\rho_m(\N)=\rho_{m0}e^{-3\N}$ and
$\rho_r(\N)=\rho_{r0}e^{-4\N}$, respectively. Defining the normalized
Hubble rate as $E(\N)\equiv H(\N)/H_0$, its evolution can therefore be
reconstructed from the radiation sector as
\begin{equation}
 E^2(\N)=\frac{\Omega_{r0}e^{-4\N}}{\Or(\N)},
 \label{eq:EfromOr}
\end{equation}
provided $\Omega_{r0}\neq0$.  Independently, the matter sector gives
\begin{equation}
 E^2(\N)=
 \frac{\Omega_{m0}e^{-3\N}}
 {1-\Ode(\N)-\Or(\N)},
\end{equation}
provided $\Omega_{m0}\neq0$.  Agreement between these two independent
reconstructions provides a useful consistency check for numerical
integrations.

The separate conservation of the effective DE component at the background
level implies
\begin{equation}
\rho_{\rm DE}'+3(1+w_{\rm DE})\rho_{\rm DE}=0.
\end{equation}
A prime denotes differentiation with respect to $\N$, except in
Appendix~\ref{app:vcdm_perturbations}, where conformal time is used.
The DE equation-of-state parameter can therefore be written as
\begin{equation}
w_{\rm DE}=-1-\frac{1}{3}
\frac{\rho_{\rm DE}'}{\rho_{\rm DE}}.
\label{eq:wmaster}
\end{equation}
The fractional energy densities satisfy
\begin{align}
\Ode'&=-\Ode\left[
3(1+w_{\rm DE})+2\frac{H'}{H}
\right],
\label{eq:masterOmegaDE}\\
\Or'&=-\Or\left(
4+2\frac{H'}{H}
\right).
\label{eq:masterOmegaR}
\end{align}
The Raychaudhuri equation then gives
\begin{equation}
2\frac{H'}{H}=-3-\Or-3w_{\rm DE}\Ode.
\label{eq:hprimegeneral}
\end{equation}
Substituting Eq.~\eqref{eq:hprimegeneral} into
Eqs.~\eqref{eq:masterOmegaDE}--\eqref{eq:masterOmegaR} yields
\begin{align}
\Ode' &= \Ode\left[
\Or-3w_{\rm DE}(1-\Ode)
\right],
\label{eq:masterClosedDE}\\
\Or' &= \Or \left[
-1+\Or+3w_{\rm DE}\Ode
\right].
\label{eq:masterClosedR}
\end{align}
Once $w_{\rm DE}$ has been expressed in terms of the chosen phase-space
variables, Eqs.~\eqref{eq:masterClosedDE}--\eqref{eq:masterClosedR} form a
closed system.  They provide the background starting point for HoloEoS in
Sec.~\ref{sec:holoeos}.

\subsection{VCDM action and parametric reconstruction}
\label{subsec:VCDMreconstruction}

We now show how the expansion history specified in
Sec.~\ref{subsec:universalBackground} can be reproduced within VCDM.
The effective DE density and pressure introduced there parametrize
the target background through the GR equations, without specifying
the DE perturbation dynamics.
VCDM reproduces this history through the potential $V(\phi)$ of a
nondynamical auxiliary scalar, without introducing an independent
DE matter field.
Pressureless matter and radiation retain their standard minimal
coupling to the metric.

We adopt the Arnowitt--Deser--Misner (ADM) decomposition
\be
 \dd s^2=-{\cal N}^2\dd t^2+\gamma_{ij}
\left(\dd x^i+{\cal N}^i\dd t\right)
 \left(\dd x^j+{\cal N}^j\dd t\right),
 \label{eq:VCDMADMmetric}
\ee
where ${\cal N}$ is the lapse, ${\cal N}^i$ is the shift vector, and
$\gamma_{ij}$ is the induced metric on constant-$t$ hypersurfaces. The
VCDM action is given by \cite{VCDMFoundation2020}
\begin{align}
 S_{\rm VCDM}
 &= \mpl^2\int \dd t\,\dd^3x\,{\cal N}\sqrt{\gamma}
 \Biggl\{\frac12\left[{}^{(3)}\!R+K_{ij}K^{ij}-K^2\right]
 \nonumber\\
 &\quad\, -V(\phi)
 -\frac{\lambda_{\rm gf}^{i}}{{\cal N}}\partial_i\phi
 -\frac34\lambda^2-\lambda(K+\phi)\Biggr\}
 \nonumber\\
 &\quad\, +S_s[g_{\mu\nu},\Psi_s].
 \label{eq:VCDMaction}
\end{align}
Here $\gamma$ is the determinant of $\gamma_{ij}$, and
${}^{(3)}\!R$ is the intrinsic Ricci scalar of the constant-$t$
hypersurfaces. The extrinsic curvature is defined by
\begin{align}
 K_{ij}&=\frac{1}{2{\cal N}}
 \left(\dot{\gamma}_{ij}-D_i{\cal N}_j-D_j{\cal N}_i\right),
 \nonumber\\
 K&=\gamma^{ij}K_{ij},
 \label{eq:VCDMextrinsiccurvature}
\end{align}
where $D_i$ denotes the covariant derivative compatible with
$\gamma_{ij}$. With the normalization in Eq.~\eqref{eq:VCDMaction},
$\phi$ and $\lambda$ have the dimensions of $H$, while $V$ has the
dimensions of $H^2$. The scalar $\phi$ is nondynamical, $\lambda$ is an auxiliary
field, and $\lambda_{\rm gf}^i$ is a vector Lagrange multiplier enforcing the
spatial-homogeneity constraint $\partial_i\phi=0$. The action
$S_s[g_{\mu\nu},\Psi_s]$ describes the standard pressureless-matter and
radiation sectors, whose fields are collectively denoted by $\Psi_s$; both
sectors are minimally coupled to $g_{\mu\nu}$.

For the background variation, we temporarily retain a general homogeneous
lapse, so that $t$ is an arbitrary time coordinate, and take
\begin{align}
 {\cal N}&={\cal N}(t),\qquad {\cal N}^i=0,
 \nonumber\\
 \gamma_{ij}&=a^2(t)\delta_{ij},\qquad
 \phi=\phi(t),\qquad \lambda=\lambda(t).
 \label{eq:VCDMFLRWansatz}
\end{align}
For this ansatz, ${}^{(3)}\!R=0$,
$K_{ij}=[\dot a/({\cal N}a)]\gamma_{ij}$, and
$K=3\dot a/({\cal N}a)$. The gauge-fixing term vanishes on the homogeneous
background, and Eq.~\eqref{eq:VCDMaction} reduces to
\begin{align}
 S_{\rm VCDM}^{(0)}
 &= \int \dd t\,\dd^3x\,{\cal N}a^3\mpl^2
 \Biggl[-3\left(\frac{\dot a}{{\cal N}a}\right)^2
 -V(\phi)\nonumber\\
 &\hspace{10mm}
 -\frac{3}{4}\lambda^2 
 -\lambda\left(3\frac{\dot a}{{\cal N}a}+\phi\right)\Biggr]
 +S_s^{(0)} .
 \label{eq:VCDMbackgroundaction}
\end{align}
For the standard matter and radiation sectors, the background action satisfies
$\delta S_s^{(0)}/\delta{\cal N}=-a^3\rho_s$ and
$\delta S_s^{(0)}/\delta a=3{\cal N}a^2P_s$, where their total energy density
and pressure are defined by
\begin{equation}
 \rho_s\equiv\rho_m+\rho_r,\qquad
 P_s\equiv P_m+P_r=\frac{\rho_r}{3}.
 \label{eq:standardMatterDensityPressure}
\end{equation}
Varying Eq.~\eqref{eq:VCDMbackgroundaction} with respect to $\lambda$ and
$\phi$, respectively, gives
\begin{equation}
\lambda=-2\frac{\dot a}{{\cal N}a}-\frac{2}{3}\phi,
\qquad \lambda=-V_{,\phi}\,,
\label{eq:VCDMlambdaequations}
\end{equation}
where $V_{,\phi}\equiv\dd V/\dd\phi$. 
Combining these two relations yields
\begin{equation}
\phi=\frac{3}{2}V_{,\phi}-3\frac{\dot a}{{\cal N}a}.
\label{eq:VCDMphiconstraint}
\end{equation}
Varying the action in Eq.~\eqref{eq:VCDMbackgroundaction}
with respect to the lapse ${\cal N}$ gives
\begin{equation}
 3\left(\frac{\dot a}{{\cal N}a}\right)^2
 -V-\frac{3}{4}\lambda^2-\lambda\phi
 =\frac{\rho_s}{\mpl^2}.
 \label{eq:VCDMlapseequation}
\end{equation}
Eliminating $\lambda$ with Eq.~\eqref{eq:VCDMlambdaequations} gives the
background constraint
\begin{equation}
 \phi^2=3V+\frac{3\rho_s}{\mpl^2}.
 \label{eq:VCDMbackgroundconstraint}
\end{equation}
Minimal coupling implies the standard conservation equations
\begin{equation}
 \frac{\dot\rho_i}{{\cal N}}
 +3\frac{\dot a}{{\cal N}a}(\rho_i+P_i)=0,
 \qquad i=m,r.
 \label{eq:VCDMstandardconservation}
\end{equation}
On an expanding background, taking the time derivative of
Eq.~\eqref{eq:VCDMbackgroundconstraint} and using
Eqs.~\eqref{eq:VCDMphiconstraint} and
\eqref{eq:VCDMstandardconservation} gives
\be
 \frac{\dot\phi}{{\cal N}}
 =\frac{3}{2}\frac{\rho_s+P_s}{\mpl^2}\,.
 \label{eq:VCDMphidot}
\ee

We now return to the cosmic-time gauge of
Sec.~\ref{subsec:universalBackground} by setting ${\cal N}=1$. Using the
Hubble expansion rate defined there and changing to the e-fold 
variable $\N=\ln a$, Eqs.~\eqref{eq:VCDMbackgroundconstraint} and
\eqref{eq:VCDMphidot} yield
\begin{align}
 \phi'&=\frac{3}{2}H\left(3\Om+4\Or\right),
 \label{eq:phiVCDMreconstruction}\\
 V\bigl(\phi(\N)\bigr)
 &=\frac{\phi^2(\N)}{3}-3H^2(\N)(\Om+\Or).
 \label{eq:VVCDMreconstruction}
\end{align}
Equations~\eqref{eq:phiVCDMreconstruction} and
\eqref{eq:VVCDMreconstruction} determine $\phi(\N)$ and $V(\N)$ and hence
give $V(\phi)$ parametrically. If $H>0$ and $\rho_s+P_s>0$, then
$\phi(\N)$ is monotonic and can be inverted to obtain a single-valued
$V(\phi)$. Conversely, a prescribed potential satisfies
\begin{align}
   H&=\frac12 V_{,\phi}-\frac13\phi,\nonumber\\
   \rho_{\rm DE}&=\mpl^2\left(
   V-\phi V_{,\phi}+\frac34 V_{,\phi}^2\right).
   \label{eq:VCDMconverse}
\end{align}
The first relation follows directly from
Eq.~\eqref{eq:VCDMphiconstraint}.
The second follows from the definition
$3\mpl^2H^2=\rho_s+\rho_{\rm DE}$ together with
Eq.~\eqref{eq:VCDMbackgroundconstraint}.
These relations reproduce the prescribed FLRW background within VCDM.
The effective DE pressure $P_{\rm DE}$ is defined by
\begin{equation}
 -2\mpl^2\dot H
 =\rho_s+P_s+\rho_{\rm DE}+P_{\rm DE}.
 \label{eq:effective_raychaudhuri}
\end{equation}
Combining this relation with the Friedmann equation and the
conservation equations for the standard species yields the effective
DE conservation law and hence Eq.~\eqref{eq:wmaster}.
This reconstruction introduces no additional degree of freedom
associated with a holographic cutoff or horizon variable.

The reconstruction extends directly to thermal massive neutrinos. With
$\rho_s=\sum_I\rho_I$ and $P_s=\sum_I P_I$, where the sums contain all
standard species and exclude the effective DE sector, its exact form is
\begin{align}
 \phi'&=\frac{3(\rho_s+P_s)}{2\mpl^2H},\\
 V\bigl(\phi(\N)\bigr)&=\frac{\phi^2(\N)}{3}
 -\frac{\rho_s(\N)}{\mpl^2}.
 \label{eq:general_species_reconstruction}
\end{align}
Equations~\eqref{eq:phiVCDMreconstruction} and
\eqref{eq:VVCDMreconstruction} are the dust-plus-radiation special case.
The massive-neutrino prescription adopted in the observational
analysis is specified in Sec.~\ref{subsec:run_settings}.

\section{Hubble-running equation of state}
\label{sec:holoeos}

\subsection{Motivation and definition}
\label{subsec:holoeos_motivation}

The representative constructions collected in
Appendix~\ref{app:representativeHDE} motivate a dependence of the effective DE sector on the expansion scale. 
Standard HDE relates the DE density to an infrared cutoff length, while local Ricci and Granda--Oliveros models express it as a linear combination of $H^2$ and 
$\dot H$. Entropy-deformed models with a Hubble-scale cutoff can yield densities proportional to nonstandard powers of $H$.
These density prescriptions motivate considering a Hubble-dependent
equation of state.
Here, we adopt the phenomenological power-law ansatz
\begin{equation}
w_{\rm DE}=w_0\left(\frac{H}{H_0}\right)^q=w_0E^q,
\label{eq:HubbleRunningEoS}
\end{equation}
where $w_0\equiv w_{\rm DE}(z=0)$ is the present-day equation-of-state
parameter and $q$ is a constant dimensionless exponent controlling
its evolution.
We refer to this two-parameter model as HoloEoS to reflect its
holographic motivation, without assuming that
Eq.~\eqref{eq:HubbleRunningEoS} follows from a specific UV/IR bound
or entropy functional.
While the CPL parametrization describes the equation of state
as a linear function of the scale factor, HoloEoS expresses it
as a power law of the Hubble rate.
Hubble-dependent equations of state have been explored in more
general settings~\cite{NojiriInhomogeneous2005,Nojiri:2005pu,Nojiri:2017opc}.
Here, we focus on realizing the specific ansatz
in Eq.~\eqref{eq:HubbleRunningEoS} within VCDM and constraining
its parameters with observational data.
The limit $q=0$ gives a DE model with constant $w_{\rm DE}$, with
$(w_0,q)=(-1,0)$ reproducing the $\Lambda$CDM expansion history.

The HoloEoS relation is imposed only on the homogeneous 
FLRW background, for which $K=3H$ in our convention.
The resulting expansion history is used to reconstruct the potential
$V(\phi)$ of the nondynamical auxiliary scalar, without introducing
any additional explicit dependence on the extrinsic curvature
into the VCDM action.
Cosmological perturbations are then derived from the reconstructed
action, rather than by promoting
Eq.~\eqref{eq:HubbleRunningEoS} to a local relation involving $K$.

\subsection{Background evolution and phantom-divide crossing}
\label{subsec:holoeos_background}

We consider the separately conserved pressureless-matter and
massless-radiation sectors introduced in
Sec.~\ref{subsec:universalBackground}.
With the HoloEoS ansatz~\eqref{eq:HubbleRunningEoS},
the background evolution equations take the form
\begin{align}
   \Ode'&=\Ode\left[\Or-3w_{\rm DE}(1-\Ode)\right],
   \label{eq:phenomOmegaDE}\\
   \Or'&=\Or\left[-1+\Or+3w_{\rm DE}\Ode\right],
   \label{eq:phenomOmegaR}\\
   E'&=-\frac{E}{2}\left(3+\Or+3w_{\rm DE}\Ode\right),
   \label{eq:holoeosE}
\end{align}
where $w_{\rm DE}=w_0E^q$.
At $\N=0$, the initial conditions are
$E=1$, $\Or=\Omega_{r0}$, and
$\Ode=1-\Omega_{m0}-\Omega_{r0}$.
The physical domain is $E>0$, $\Ode\geq0$, $\Or\geq0$, and
$\Om=1-\Ode-\Or\geq0$.
The first two equations also form a closed, generally non-autonomous,
two-dimensional system if $E$ is eliminated using Eq.~\eqref{eq:EfromOr}.
If $E$ is evolved explicitly, its equation preserves this same
radiation normalization; the three initial values cannot be chosen
independently of the present density fractions.

For a general conserved standard sector, including thermal massive
neutrinos, the equivalent background formulation is
\begin{align}
 \frac{\rho_{\rm DE}'}{\rho_{\rm DE}}
 &=-3(1+w_0E^q),\nonumber\\
 E^2&=\frac{\rho_s+\rho_{\rm DE}}{3\mpl^2H_0^2},
 \label{eq:holoeos_general_background}
\end{align}
where $\rho_s$ includes the thermal neutrino density through its
relativistic-to-nonrelativistic transition. The dust-plus-radiation
system above follows when $\rho_s=\rho_{m0}a^{-3}+\rho_{r0}a^{-4}$.

For $w_0<0$ and $q\neq0$, the condition $w_{\rm DE}=-1$
selects the following value of the normalized Hubble rate $E=H/H_0$:
\begin{equation}
E_\times=\left(-\frac{1}{w_0}\right)^{1/q}.
\label{eq:HubbleRunningCrossing}
\end{equation}
A phantom-divide crossing occurs only if the background trajectory reaches
$E=E_\times$ and $w_{\rm DE}+1$ changes sign.
Using $w_{\rm DE}'=qw_{\rm DE}H'/H$ together with
the Raychaudhuri equation, we obtain
\begin{equation}
\left.w_{\rm DE}'\right|_\times
=\frac{q(\rho_s+P_s)_\times}{2\mpl^2H_\times^2}
=\frac{q}{2}\left(3\Om+4\Or\right)_\times.
\label{eq:holoeos_crossing_slope}
\end{equation}
The final equality assumes the dust-plus-radiation background.
The expression in terms of $\rho_s+P_s$ remains valid in the
presence of thermal massive neutrinos, provided that $\rho_s$
and $P_s$ include all standard species.
For $q>0$ and positive standard-species enthalpy, any regular
crossing is therefore simple and proceeds from $w_{\rm DE}<-1$
to $w_{\rm DE}>-1$ as the Universe expands.
For $q>0$ and $-1<w_0<0$, one has $E_\times>1$.
A regular trajectory evolving from an early radiation era with
$E>E_\times$ to the present value $E=1$ necessarily crosses
the phantom divide, since $w_{\rm DE}=w_0E^q<-1$ in that era
and $w_{\rm DE}=w_0>-1$ today.
This conclusion follows from continuity and does not require $H$
to be monotonic throughout the intervening evolution.
For $q=0$, the equation of state is constant and no crossing occurs;
in particular, $w_0=-1$ gives $w_{\rm DE}\equiv-1$.

The effective background density and pressure are
\begin{equation}
\rho_{\rm DE}=3\mpl^2H^2\Ode,\qquad
P_{\rm DE}=w_{\rm DE}\rho_{\rm DE}.
\end{equation}
Differentiating the second relation and using separate DE conservation
yields the following expression, whose second line assumes the
dust-plus-radiation background:
\begin{align}
P_{\rm DE}'
&=\rho_{\rm DE}\left[w_{\rm DE}'-3w_{\rm DE}(1+w_{\rm DE})\right]\nonumber\\
&=-P_{\rm DE}\Biggl[3(1+w_{\rm DE})+\frac{q}{2}
   \left(3+\Or+3w_{\rm DE}\Ode\right)\Biggr].
   \label{eq:holoeosPressureDerivative}
\end{align}
At the crossing, $\rho_{\rm DE}'=0$ whereas
$P_{\rm DE}'=\rho_{\rm DE}w_{\rm DE}'$ is finite.
For $q>0$ and $\rho_{\rm DE}>0$, the density has a local maximum,
since $\rho_{\rm DE}''|_\times
=-3\rho_{{\rm DE},\times}w_{\rm DE}'|_\times<0$.

\subsection{Reconstructed VCDM potential}
\label{subsec:holoeos_potential}

Solving the background equations fixes $H(\N)$, $\Om(\N)$, and
$\Or(\N)$. The VCDM reconstruction then gives
\begin{align}
   \phi(\N)&=\phi_0+
   \frac{3}{2}\int_0^{\N}\dd\widetilde\N\,
   H(\widetilde\N)
   \left[3\Om(\widetilde\N)+4\Or(\widetilde\N)\right],
   \label{eq:holoeosPhiIntegral}\\
   V\bigl(\phi(\N)\bigr)
   &=\frac{\phi^2(\N)}{3}
   -3H^2(\N)\left[\Om(\N)+\Or(\N)\right].
   \label{eq:holoeosPotential}
\end{align}
Here $\widetilde\N$ is an integration variable and $\phi_0$ fixes
the auxiliary-field origin, with a corresponding change in the
representative potential. It is not an additional cosmological
parameter. For $H>0$ and $\rho_s+P_s>0$, the field satisfies
$\phi'>0$, so the parametric curve defines a single-valued
$V(\phi)$ on the field interval traversed by the background.
Its extension outside that interval is not fixed by this reconstruction.
When thermal massive neutrinos are included, the same construction
applies, with the dust-plus-radiation expressions above replaced
by the exact sums over standard species in
Eq.~\eqref{eq:general_species_reconstruction}.

Each point in cosmological parameter space determines a reconstructed
potential along the corresponding background trajectory, so no
additional arbitrary function is sampled.
A Boltzmann solver can evaluate the required background combinations
directly, without storing a global interpolation of $V(\phi)$.
The likelihood analysis therefore constrains a family of reconstructed
VCDM theories, rather than a single fixed potential common to
all parameter points.

The crossing has a direct interpretation in the potential.
Differentiating the first relation in Eq.~\eqref{eq:VCDMconverse}
and using Eq.~\eqref{eq:VCDMphidot} gives
\begin{equation}
\dot H=\frac{\rho_s+P_s}{4\mpl^2}\left(3V_{,\phi\phi}-2\right).
 \label{eq:holoeos_Hdot_potential}
\end{equation}
Comparison with Eq.~\eqref{eq:effective_raychaudhuri} yields
\begin{equation}
 V_{,\phi\phi}
 =-\frac{2(1+w_{\rm DE})\rho_{\rm DE}}{3(\rho_s+P_s)}.
 \label{eq:potential_curvature_crossing}
\end{equation}
These relations agree with Ref.~\cite{VCDMFoundation2020} and hold
also with thermal massive neutrinos.
For $\rho_{\rm DE}>0$ and $\rho_s+P_s>0$, a simple phantom-divide
crossing is an inflection of $V(\phi)$.
On the $q>0$ branch, $V_{,\phi\phi}$ changes from positive to
negative while $\dot\phi$ stays positive.
For a smooth background with finite nonzero $H$ and
$\rho_s+P_s$, both the reconstruction and its field derivatives are
finite at this crossing. 
In the $\Lambda$CDM limit, $V_{,\phi\phi}=0$ along the entire
background trajectory, so the reconstructed potential takes
the affine form $V(\phi)=A\phi+B$, with constant coefficients
$A$ and $B$, over the corresponding field range.
The vanishing second derivative therefore does not indicate
an isolated inflection point.

\subsection{Perturbative interpretation and validity domain}
\label{subsec:holoeos_validity}

For a separately conserved effective DE component, we define the
background adiabatic ratio wherever $\rho_{\rm DE}'\neq0$ as
\begin{align}
c_{a,\rm DE}^2
&\equiv\frac{P_{\rm DE}'}{\rho_{\rm DE}'}
=w_{\rm DE}-\frac{w_{\rm DE}'}{3(1+w_{\rm DE})}\nonumber\\
&=w_{\rm DE}+\frac{qw_{\rm DE}}{6(1+w_{\rm DE})}
 \left(3+\Or+3w_{\rm DE}\Ode\right).
\label{eq:adiabaticSoundSpeed}
\end{align}
The last equality uses the dust-plus-radiation background.
At a simple phantom-divide crossing, $\rho_{\rm DE}'$ vanishes,
whereas $P_{\rm DE}'$ remains finite and nonzero, causing this
ratio to diverge.
This divergence arises because the background DE density ceases
to provide a valid local time variable at the crossing;
it does not by itself imply a singular physical propagation speed.
A purely adiabatic fluid closure,
$\delta P_{\rm DE}=c_{a,\rm DE}^2\delta\rho_{\rm DE}$,
would therefore be problematic at the
crossing~\cite{KunzSapone:2006}.
In VCDM, the metric and standard-matter perturbations are instead
governed by the field equations, without introducing an independent
DE matter field or assigning it a sound speed.
The momentum constraint given in
Appendix~\ref{app:vcdm_perturbations} contains no denominator
proportional to $1+w_{\rm DE}$ and remains finite at the crossing
for modes with comoving wavenumber $k>0$, provided that the
standard-species enthalpy is positive.
The usual stability conditions for the standard-matter sector
must still be satisfied, but the divergent effective adiabatic
ratio does not constitute an additional scalar stability
condition in VCDM~\cite{VCDMFoundation2020,VCDMArora2025}.

The early-time behavior can also be examined analytically in the
dust-plus-radiation approximation of
Sec.~\ref{subsec:universalBackground}.
For $w_0<0$ and $q>0$, the radiation-dominated branch satisfies
$E(z)\simeq\sqrt{\Omega_{r0}}(1+z)^2$, so that
\begin{equation}
w_{\rm DE}(z)\simeq
w_0\Omega_{r0}^{q/2}(1+z)^{2q}.
\label{eq:w_asymptotic_early}
\end{equation}
The conservation equation for the effective DE component gives
\begin{equation}
\frac{\dd\ln\rho_{\rm DE}}{\dd\ln(1+z)}
=3\left[1+w_{\rm DE}(z)\right].
\end{equation}
For a reference redshift $z_\star$ during radiation domination,
integration using the leading radiation-dominated approximation
yields
\begin{eqnarray}
\hspace{-0.7cm}
\ln\frac{\rho_{\rm DE}(z)}{\rho_{\rm DE}(z_\star)}
&\simeq& 3\ln\frac{1+z}{1+z_\star} \nonumber \\
\hspace{-0.7cm}
&& +\frac{3w_0\Omega_{r0}^{q/2}}{2q}
\bigl[(1+z)^{2q}-(1+z_\star)^{2q}\bigr].
\label{eq:rho_superexponential}
\end{eqnarray}
For $w_0<0$ and $q>0$, the negative term proportional to
$(1+z)^{2q}$ dominates the logarithmic term as $z\to\infty$.
Consequently, both $\rho_{\rm DE}$ and
$|P_{\rm DE}|=|w_{\rm DE}|\rho_{\rm DE}$ vanish faster than
any inverse power of $1+z$, even though $w_{\rm DE}\to-\infty$.
Their contributions relative to radiation therefore vanish,
making the assumed radiation-dominated asymptotics self-consistent.
With thermal massive neutrinos, the early radiation normalization
includes their relativistic contribution; the same suppression of
the DE density and pressure follows.
Equation~(\ref{eq:rho_superexponential}) gives the leading
asymptotic behavior of $\ln\rho_{\rm DE}$; it is not a uniformly
accurate expression for the density through matter--radiation
equality.

We sample $q$ over the prior range $[0,2]$ and $w_0$ over $[-2,0]$.
At the boundary $q=0$, the equation of state is constant,
giving the exact power-law evolution
$\rho_{\rm DE}(z)/\rho_{{\rm DE},0}=(1+z)^{3(1+w_0)}$.
This differs from the $q>0$, $w_0<0$ case, in which the density
vanishes faster than any inverse power of $1+z$ as $z\to\infty$.
For $w_0\leq0$, the ratio $\rho_{\rm DE}/\rho_r$ nevertheless
vanishes as $z\to\infty$.
At the boundary $w_0=0$, one has $w_{\rm DE}=0$ for any finite
$E$, so the effective DE component behaves as pressureless matter
at the background level.
Negative values of $q$, for which $w_{\rm DE}\to0$ as
$H\to\infty$, are excluded from the sampled prior.

\section{Observational constraints}
\label{sec:constraints}

In this section, we constrain HoloEoS with observational data and
compare it with flat $\Lambda$CDM and CPL-VCDM.
We first specify the data and likelihoods in
Sec.~\ref{subsec:datasets}, followed by the models, priors, and
numerical implementation in Sec.~\ref{subsec:run_settings}.
For each data combination, PolyChord supplies the posterior samples,
while separate MH simulated-annealing searches
determine the best-fit likelihoods used for the AIC.

\subsection{Data and likelihood specification}
\label{subsec:datasets}

We consider two data combinations:
\begin{align}
\mathcal D_0&=\mathrm{CMB\text{-}SPA}
+\mathrm{DESI\ DR2}_{\rm updated},\nonumber\\
\mathcal D_1&=\mathcal D_0+\mathrm{DESY5\ Dovekie}.
\label{eq:data_combinations}
\end{align}
Here CMB-SPA denotes the combined SPT, Planck, and ACT CMB data,
including the primary-anisotropy and lensing likelihoods specified below.
These likelihoods are evaluated directly, rather than replaced by
compressed CMB distance priors.
The same optical-depth and calibration information is included in
both data combinations.

\paragraph{Primary CMB anisotropies.}
We denote temperature and E-mode polarization auto-spectra by TT
and EE, respectively, and their cross-spectrum by TE.
We use the Planck 2018 Commander low-$\ell$ TT likelihood for multipoles
$2\leq\ell\leq29$, implemented using the released native spline data,
together with the foreground-marginalized Plik-lite TT, TE, and EE
likelihood~\cite{PlanckLikelihood2018}.
For the high-$\ell$ Planck contribution, we retain TT bins up to
$\ell=1000$ and TE/EE bins up to $\ell=600$.
These multipole cuts follow the SPA prescription adopted for combining
Planck with the ACT DR6 CMB-only TT, TE, and EE
likelihood~\cite{ACTDR6CMB}.
We also include the SPT-3G D1 lite temperature and polarization
likelihood~\cite{SPT3GD1}, evaluated through \texttt{candl}.
The Planck and ACT temperature calibrations share a sampled parameter,
$A_{\rm act}=A_{\rm planck}$, while ACT has a separate polarization
calibration parameter, $P_{\rm act}$.
The ACT calibration factors are applied to the spectra once within
the ACT likelihood interface.
The SPT temperature and polarization calibration parameters are
$T_{\rm cal}$ and $E_{\rm cal}$, respectively.

\paragraph{CMB lensing.}
We adopt the joint ACT DR6+Planck baseline lensing likelihood,
including the response corrections and Hartlap correction implemented
in the released likelihood
adapter~\cite{ACTDR6Lensing,ACTDR6Cosmology,PlanckPR4Lensing}.
No separate Planck lensing contribution is added.
We also include the CMB lensing-potential power-spectrum
($C_L^{\phi\phi}$) component of the SPT-3G MUSE
likelihood~\cite{SPTMUSE}, while excluding its delensed EE component
from the primary CMB data combination.
Here $L$ is the lensing multipole, and $\phi$ in
$C_L^{\phi\phi}$ denotes the CMB lensing potential, distinct from
the VCDM auxiliary field.

\paragraph{Optical depth and calibration information.}
We adopt the following Gaussian priors:
\begin{equation}
\tau \sim\mathcal N(0.051,0.006^2),\qquad
T_{\rm cal} \sim\mathcal N(1,0.0036^2).
\label{eq:external_gaussians}
\end{equation}
Here $\tau$ is the Thomson optical depth due to reionization,
and $T_{\rm cal}$ is the dimensionless calibration parameter
for the SPT-3G temperature amplitude.
The notation $\mathcal N(\mu,\sigma^2)$ denotes a Gaussian
distribution with mean $\mu$ and variance $\sigma^2$.
Both normalized Gaussian priors are implemented directly
in the PolyChord prior transform.
No SimAll, SROLL, or separate $\tau$ likelihood is included.
The internal SPT priors are disabled, and the ACT likelihood interface
applies no additional Gaussian prior to its temperature calibration.
Thus each external Gaussian prior enters the inference once.
The remaining parameter priors are specified in
Sec.~\ref{subsec:run_settings}.

\paragraph{Updated DESI DR2 distances.}
We use the DESI DR2 galaxy and quasar BAO distance measurements
and their covariance~\cite{DESI:2025DR2BAO}, together with the
updated DR2 Ly$\alpha$ distance constraints from the joint BAO
and Alcock--Paczy\'nski full-shape analysis at an effective
redshift $z_{\rm eff}=2.33$~\cite{DESILyAFS2026}.
The updated Ly$\alpha$ distance pair replaces the original BAO pair,
which is not included separately.
For the spatially flat background considered here, the transverse
comoving distance $D_M$, Hubble distance $D_H$, and volume-averaged
distance $D_V$ are defined by
\begin{align}
D_M(z)&=c\int_0^z\frac{\dd z'}{H(z')},\\
D_H(z)&=\frac{c}{H(z)},\\
D_V(z)&=\left[zD_M^2(z)D_H(z)\right]^{1/3},
\end{align}
where $c$ is the speed of light.
We denote the comoving sound horizon at the baryon drag epoch by $r_d$.
The observables are $D_M/r_d$, $D_H/r_d$, and $D_V/r_d$, depending
on the tracer.
Only the compressed Ly$\alpha$ distance constraints enter the analysis;
no galaxy power-spectrum full-shape likelihood is evaluated.

\paragraph{DESY5 with Dovekie.}
We also analyze the DES five-year supernova
compilation~\cite{DES:2024tys} with the Dovekie recalibration and
the associated covariance matrix including statistical and systematic
uncertainties~\cite{Dovekie2025}.
Adding this sample to the baseline combination $\mathcal D_0$
defines $\mathcal D_1$.
The supernova sample contains 1820 objects after restricting to
positive redshifts and includes the low-redshift sample.
We analytically marginalize over the common absolute-magnitude offset.
No supernova absolute-distance calibration or local $H_0$ likelihood
is added.

The primary CMB spectra and CMB lensing constrain the early-time
sound horizon and the growth of structure.
BAO measurements constrain radial and transverse distances in units
of $r_d$, while supernovae constrain the relative distance--redshift
relation at low redshifts.
For each data combination $\mathcal D_i$ ($i=0,1$), all three models
are evaluated with identical data files and likelihood settings.

\subsection{Models, neutrinos, priors, and numerical conventions}
\label{subsec:run_settings}

\paragraph{Models.}
The CPL-VCDM model realizes the background equation of state
$w_{\rm DE}(a)=w_0+w_a(1-a)$ within
VCDM~\cite{VCDMFoundation2020,VCDMArora2025}.
For both HoloEoS and CPL-VCDM, each choice of background parameters
specifies a reconstructed potential $V(\phi)$, and the perturbations
follow from the corresponding action.
HoloEoS varies $(w_0,q)$, whereas CPL-VCDM varies $(w_0,w_a)$.
For the $\Lambda$CDM benchmark, $w_{\rm DE}=-1$ is fixed.
No independent DE perturbation parameters are introduced.
All runs employ the VCDM perturbation equations written in terms of
Newtonian-gauge-type Bardeen variables, as explained in
Appendix~\ref{app:vcdm_perturbations}. In the $\Lambda$CDM limit,
these equations recover the standard GR perturbation dynamics.

\paragraph{Neutrinos and other fixed prescriptions.}
We adopt the same massive-neutrino sector in all six runs:
three thermal species with equal masses
$m_{\nu,i}=0.02\,\mathrm{eV}$ ($i=1,2,3$), giving a total mass
$\sum_i m_{\nu,i}=0.06\,\mathrm{eV}$, and no additional massless
neutrino species.
We set $T_\nu/T_\gamma=0.7163687246184776$, where $T_\nu$ and
$T_\gamma$ denote the neutrino and photon temperatures, respectively.
In the relativistic limit, this prescription corresponds to
$N_{\rm eff}=3.044$, where $N_{\rm eff}$ is the effective number
of relativistic neutrino species.
The CLASS background retains the thermal neutrino density
$\rho_\nu(z)$ and pressure $P_\nu(z)$ through the continuous
transition from relativistic to nonrelativistic behavior.
For this species content, the HoloEoS background and VCDM reconstruction
are expressed through the total standard-species density and pressure
in Eqs.~\eqref{eq:holoeos_general_background} and
\eqref{eq:general_species_reconstruction}.
The constraints are conditional on this fixed degenerate-mass
prescription, which does not reproduce the mass splittings inferred
from neutrino oscillations.

The primordial helium mass fraction $Y_{\rm He}$ is determined
using the same PArthENoPE-based big bang nucleosynthesis (BBN)
interpolation table in all models.
Nonlinear corrections are modeled with HMcode2020~\cite{HMcode2020},
with density and pressure inputs from the VCDM implementation and
numerical domain checks.
Its application here is a modeling approximation, rather than a
dedicated calibration of nonlinear structure formation in VCDM.

\paragraph{Priors.}
Table~\ref{tab:priors} summarizes the priors adopted for both data
combinations.
The six common cosmological parameters and four calibration nuisance
parameters have identical priors in all three models, giving ten
varying parameters in $\Lambda$CDM and twelve in either extension.
Here $\omega_b\equiv\Omega_{b0}h^2$ and
$\omega_{\rm cdm}\equiv\Omega_{{\rm cdm},0}h^2$ are the present-day physical
baryon and CDM density parameters, while $A_s$ and $n_s$ are the
amplitude and spectral index of the primordial scalar power spectrum.
We sample $H_0$ directly.
For CPL-VCDM, the joint prior is uniform on
\begin{equation}
\begin{split}
\mathcal R_{\rm CPL}
=\bigl\{(w_0,w_a)\in[-3,1]\times[-3,2]:\\
w_0+w_a<0\bigr\},
\end{split}
\label{eq:cpl_prior_domain}
\end{equation}
whose area is $15.5$.
This domain is sampled directly through the PolyChord prior transform.
The restriction $w_0+w_a<0$ makes the effective DE density negligible
relative to baryons and CDM as $a\to0$; it is a prior on the early-time
background behavior, rather than a VCDM no-ghost condition.
For HoloEoS, the joint prior is uniform on
$(w_0,q)\in[-2,0]\times[0,2]$, whose area is $4$.
These domains specify the posterior priors; neither is a VCDM
stability condition.

%%%%%%%%%%%%%%%%%%%%%%%%%%%%%%%%%%%%
\begin{table}[!tbp]
\centering
\caption{Priors adopted for both data combinations.
$\mathcal U$ denotes a normalized uniform distribution over the
specified interval or joint domain, and $\mathcal N$ denotes
a normalized Gaussian distribution.
The CPL entry refers to the joint prior defined in
Eq.~(\ref{eq:cpl_prior_domain}); its one-dimensional marginal
distributions are not uniform.
Only parameters included in a given model are sampled.}
\label{tab:priors}
\begin{ruledtabular}
\begin{tabular}{lc}
Parameter & Prior \\
\hline
$\omega_b$ & $\mathcal U(0.005,0.1)$ \\
$\omega_{\rm cdm}$ & $\mathcal U(0.001,0.99)$ \\
$H_0\ [\mathrm{km\,s^{-1}\,Mpc^{-1}}]$ & $\mathcal U(40,100)$ \\
$\ln(10^{10}A_s)$ & $\mathcal U(1.61,3.91)$ \\
$n_s$ & $\mathcal U(0.8,1.2)$ \\
$\tau$ & $\mathcal N(0.051,0.006^2)$ \\
\hline
HoloEoS $w_0$ & $\mathcal U(-2,0)$ \\
HoloEoS $q$ & $\mathcal U(0,2)$ \\
CPL-VCDM $(w_0,w_a)$ & $\mathcal U(\mathcal R_{\rm CPL})$ \\
\hline
$A_{\rm planck}=A_{\rm act}$ & $\mathcal U(0.5,1.5)$ \\
$P_{\rm act}$ & $\mathcal U(0.9,1.1)$ \\
$T_{\rm cal}$ & $\mathcal N(1,0.0036^2)$ \\
$E_{\rm cal}$ & $\mathcal U(0.8,1.2)$ \\
\end{tabular}
\end{ruledtabular}
\end{table}
%%%%%%%%%%%%%%%%%%%%%%%%%%%%%%%%%%

\paragraph{Numerical implementation and posterior sampling.}
The analysis is performed with
\texttt{MontePython}~\cite{Audren:2012wb,Brinckmann:2018cvx},
PolyChord~\cite{Handley:2015a,Handley:2015b}, and the VCDM
implementation of \texttt{CLASS}~3.4.0~\cite{Blas:2011rf}.
The CLASS settings are $\ell_{\max}=9000$,
\texttt{accurate\_lensing=0}, \texttt{delta\_l\_max=500},
\texttt{l\_linstep=10}, and \texttt{l\_logstep=1.013}.
The background and perturbation integration tolerances are
$10^{-7}$ and $10^{-6}$, respectively.
For HoloEoS, the DE density evolution is integrated backward from
$\ln a=0$ to $\ln a=-48.4$.
We tabulate $\ln\rho_{\rm DE}$ and its derivative with respect to
$\ln a$ over 5000 intervals and apply cubic Hermite interpolation
between grid points.
The complete numerical configuration is recorded in the run
configuration files.

Each PolyChord run has $N_{\rm live}=200$ live points and an initial
prior-sampling target of 400 accepted points.
The remaining sampler settings are \texttt{num\_repeats=30},
grade repeats $[30,30]$, no clustering, a compression factor of $0.8$,
and a termination criterion of $0.01$.
Posterior summaries are computed directly from the native weighted
samples, without burn-in removal, thinning, or inclusion of duplicate
converted files.
The effective sample sizes calculated from the posterior weights
are approximately $1680$--$1753$; these values do not by themselves
establish convergence.
All quoted credible intervals are central intervals computed from
the weighted posterior. The separate simulated-annealing searches
used for the best-fit comparison in Sec.~\ref{subsec:model_comparison}
do not replace these posterior samples.

\subsection{Cosmological parameters and phantom-divide crossing}
\label{subsec:posteriors}
%

% BEGIN EMBEDDED TABLE: posteriors
\begin{table*}[!tbp]
\centering
\caption{Weighted posterior medians and central $68\%$ credible intervals.
$\mathcal D_0$ denotes CMB-SPA plus the updated DESI DR2 distances,
while $\mathcal D_1$ additionally includes DESY5 Dovekie.
$H_0$ is quoted in $\mathrm{km\,s^{-1}\,Mpc^{-1}}$.
Both DE extensions employ VCDM perturbations, 
and all three models adopt the same fixed total neutrino mass,
$\sum_i m_{\nu,i}=0.06\,\mathrm{eV}$.
$\Omega_{m0}$ denotes the present total matter density fraction,
including massive neutrinos.
The clustering parameters $\sigma_8$ and $S_8$ are evaluated at $z=0$.}
\label{tab:posteriors}
\small
\renewcommand{\arraystretch}{1.28}
\setlength{\tabcolsep}{4pt}
\begin{ruledtabular}
\begin{tabular}{lcccccc}
 & \multicolumn{3}{c}{$\mathcal D_0$: CMB-SPA + updated DESI DR2} & \multicolumn{3}{c}{$\mathcal D_1$: $\mathcal D_0$ + DESY5 Dovekie} \\
Parameter & $\Lambda$CDM & HoloEoS & CPL-VCDM & $\Lambda$CDM & HoloEoS & CPL-VCDM \\
\hline
$100\omega_b$ & $2.245^{+0.009}_{-0.009}$ & $2.240^{+0.009}_{-0.010}$ & $2.239^{+0.010}_{-0.010}$ & $2.244^{+0.009}_{-0.009}$ & $2.240^{+0.010}_{-0.009}$ & $2.241^{+0.010}_{-0.009}$ \\
$\omega_{\rm cdm}$ & $0.1181^{+0.0006}_{-0.0006}$ & $0.1197^{+0.0008}_{-0.0008}$ & $0.1198^{+0.0007}_{-0.0008}$ & $0.1184^{+0.0006}_{-0.0006}$ & $0.1196^{+0.0007}_{-0.0008}$ & $0.1195^{+0.0007}_{-0.0008}$ \\
$H_0$ & $68.13^{+0.24}_{-0.25}$ & $66.85^{+1.35}_{-1.43}$ & $64.57^{+1.90}_{-1.98}$ & $68.02^{+0.25}_{-0.25}$ & $67.53^{+0.49}_{-0.55}$ & $67.41^{+0.50}_{-0.57}$ \\
$\ln(10^{10}A_s)$ & $3.058^{+0.011}_{-0.010}$ & $3.043^{+0.011}_{-0.011}$ & $3.042^{+0.010}_{-0.010}$ & $3.056^{+0.008}_{-0.010}$ & $3.043^{+0.010}_{-0.010}$ & $3.045^{+0.009}_{-0.010}$ \\
$n_s$ & $0.9724^{+0.0028}_{-0.0027}$ & $0.9697^{+0.0030}_{-0.0030}$ & $0.9695^{+0.0029}_{-0.0032}$ & $0.9722^{+0.0029}_{-0.0031}$ & $0.9697^{+0.0030}_{-0.0031}$ & $0.9699^{+0.0030}_{-0.0031}$ \\
$\tau$ & $0.059^{+0.006}_{-0.005}$ & $0.053^{+0.006}_{-0.006}$ & $0.053^{+0.005}_{-0.005}$ & $0.058^{+0.005}_{-0.005}$ & $0.054^{+0.005}_{-0.006}$ & $0.054^{+0.005}_{-0.005}$ \\
$w_0$ & $-1$ (fixed) & $-0.833^{+0.110}_{-0.093}$ & $-0.540^{+0.225}_{-0.194}$ & $-1$ (fixed) & $-0.881^{+0.040}_{-0.038}$ & $-0.822^{+0.055}_{-0.051}$ \\
$q$ & --- & $0.545^{+0.280}_{-0.225}$ & --- & --- & $0.453^{+0.143}_{-0.133}$ & --- \\
$w_a$ & --- & --- & $-1.393^{+0.524}_{-0.639}$ & --- & --- & $-0.660^{+0.190}_{-0.205}$ \\
$\Omega_{m0}$ & $0.304^{+0.003}_{-0.003}$ & $0.319^{+0.015}_{-0.013}$ & $0.342^{+0.023}_{-0.020}$ & $0.306^{+0.003}_{-0.003}$ & $0.313^{+0.005}_{-0.005}$ & $0.314^{+0.005}_{-0.005}$ \\
$\sigma_8$ & $0.814^{+0.004}_{-0.004}$ & $0.812^{+0.012}_{-0.013}$ & $0.793^{+0.016}_{-0.018}$ & $0.814^{+0.004}_{-0.004}$ & $0.817^{+0.007}_{-0.007}$ & $0.815^{+0.007}_{-0.007}$ \\
$S_8$ & $0.821^{+0.006}_{-0.007}$ & $0.838^{+0.010}_{-0.009}$ & $0.847^{+0.011}_{-0.012}$ & $0.822^{+0.006}_{-0.006}$ & $0.835^{+0.007}_{-0.007}$ & $0.834^{+0.007}_{-0.007}$ \\
\end{tabular}
\end{ruledtabular}
\end{table*}
% END EMBEDDED TABLE: posteriors

We now examine the cosmological constraints and the inferred DE
histories.
Table~\ref{tab:posteriors} presents the marginalized constraints
as posterior medians with central $68\%$ credible intervals.
For the full data combination $\mathcal D_1$, HoloEoS gives
$w_0=-0.881^{+0.040}_{-0.038}$ and $q=0.453^{+0.143}_{-0.133}$,
whereas CPL-VCDM gives
$w_0=-0.822^{+0.055}_{-0.051}$ and
$w_a=-0.660^{+0.190}_{-0.205}$.
In both models, the posteriors favor histories that are
quintessence-like today and phantom-like at higher redshift.

\begin{figure*}[!tp]
\centering
\includegraphics[width=\textwidth]{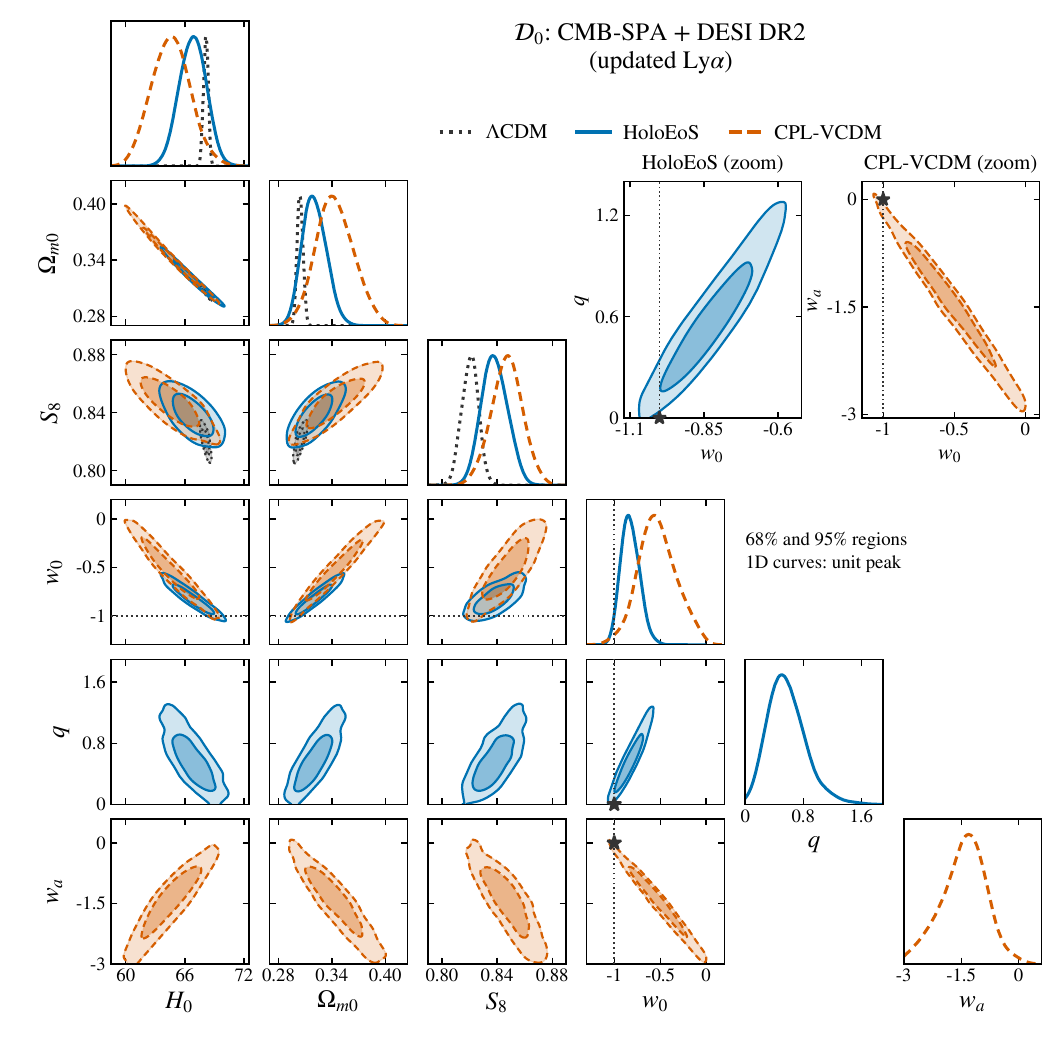}
\caption{Marginalized posteriors for $\mathcal D_0$
(CMB-SPA plus the updated DESI DR2 distances).
Gray, blue, and orange correspond to $\Lambda$CDM, HoloEoS,
and CPL-VCDM, respectively.
The diagonal densities are normalized to unit peak.
Dark and light contours enclose $68\%$ and $95\%$ of the
marginalized posterior probability, respectively.
The parameter $q$ is sampled only in HoloEoS and $w_a$ only
in CPL-VCDM, so the $(q,w_a)$ cell is left blank.
Dotted lines indicate $w_0=-1$, while stars mark the
$\Lambda$CDM limits $(w_0,q)=(-1,0)$ and $(w_0,w_a)=(-1,0)$.
No posterior densities are plotted for fixed parameters.
The upper-right insets show enlarged views of the $(w_0,q)$
and $(w_0,w_a)$ planes, using the same contours as the
corresponding main panels with separately chosen axis ranges.
$H_0$ is expressed in $\mathrm{km\,s^{-1}\,Mpc^{-1}}$.}
\label{fig:triangle_D0}
\end{figure*}

\begin{figure*}[!tp]
\centering
\includegraphics[width=\textwidth]{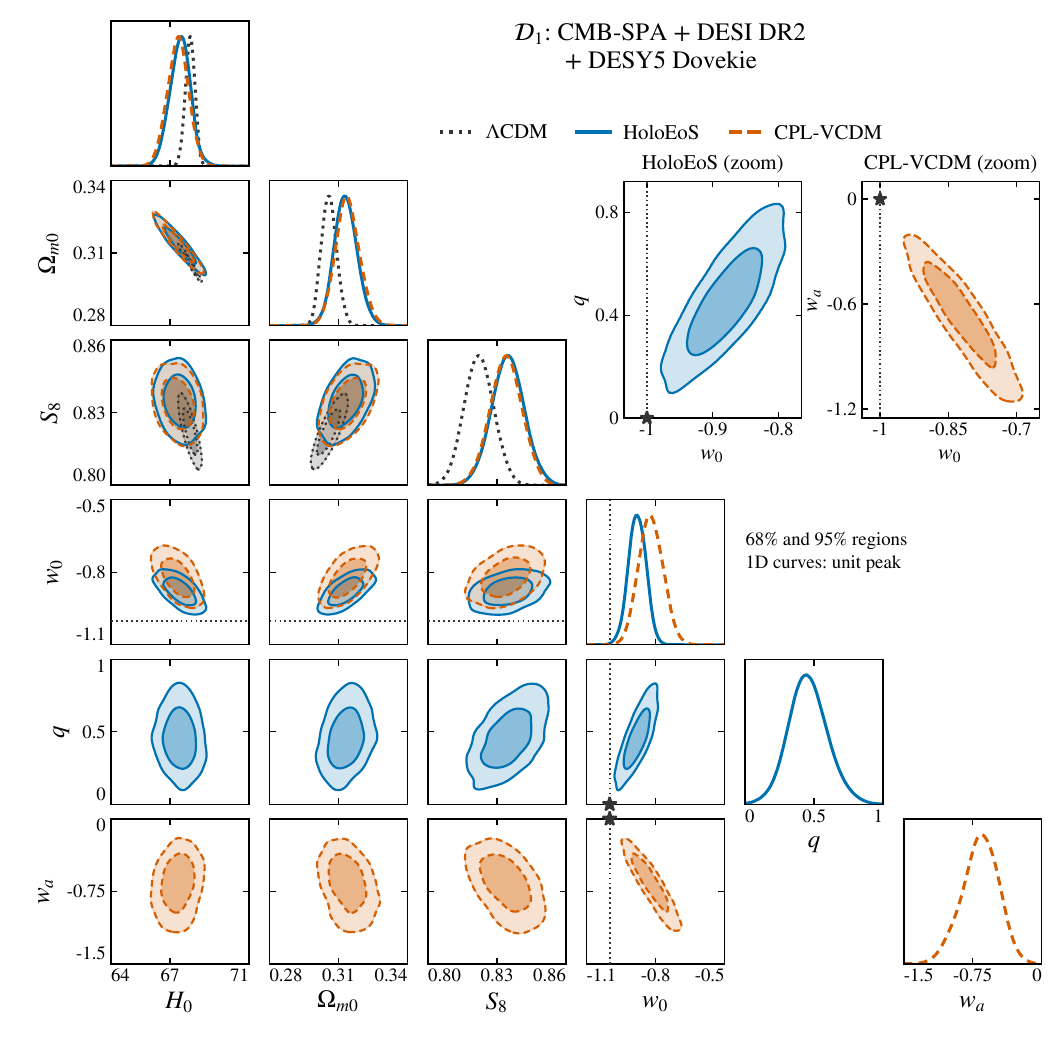}
\caption{Marginalized posteriors for $\mathcal D_1$, obtained by adding
DESY5 Dovekie to $\mathcal D_0$.
The main panels adopt narrower axis ranges than
Fig.~\ref{fig:triangle_D0} to show the tighter constraints
in greater detail; all $95\%$ contours remain fully visible.
The parameter order, colors, reference markers, and contour levels
are the same as in Fig.~\ref{fig:triangle_D0}.
The upper-right insets show enlarged views of the $(w_0,q)$
and $(w_0,w_a)$ planes for HoloEoS and CPL-VCDM, respectively.
These posteriors are conditional on each model, and contour
overlap does not quantify relative model probabilities.}
\label{fig:triangle_D1}
\end{figure*}

Figures~\ref{fig:triangle_D0} and \ref{fig:triangle_D1} present the
one- and two-dimensional marginalized posteriors for $\mathcal D_0$
and $\mathcal D_1$, respectively, obtained with
\texttt{GetDist}~\cite{GetDist2019}.
Both figures adopt the parameter order
$(H_0,\Omega_{m0},S_8,w_0,q,w_a)$ and the same model colors,
with $H_0$ expressed in $\mathrm{km\,s^{-1}\,Mpc^{-1}}$.
Figure~\ref{fig:triangle_D1} adopts narrower axis ranges than
Fig.~\ref{fig:triangle_D0} to show the more tightly constrained
posterior distributions in greater detail.
The shaded two-dimensional regions enclose $68\%$ and $95\%$
of the marginalized posterior probability; they are credible regions,
rather than profile-likelihood contours.
The Gaussian priors in Eq.~(\ref{eq:external_gaussians}) are already
accounted for in the posterior weights and are not imposed again
when generating these plots.

The panels for the common cosmological parameters show a pronounced
anticorrelation between $H_0$ and $\Omega_{m0}$.
For $\mathcal D_0$, both VCDM extensions allow broader parameter
ranges than $\Lambda$CDM, particularly CPL-VCDM.
The correlations among the DE parameters also differ:
$w_0$ and $q$ are positively correlated in HoloEoS, whereas
$w_0$ and $w_a$ are anticorrelated in CPL-VCDM.
In some $\mathcal D_0$ projections, the CPL-VCDM contours reach
the lower prior bound $w_a=-3$, indicating that this tail is
truncated by the adopted prior.

Adding DESY5 Dovekie to $\mathcal D_0$ narrows the allowed parameter
regions and brings the HoloEoS and CPL-VCDM posteriors into closer
agreement in the $(H_0,\Omega_{m0},S_8)$ projections.
The parameters $q$ and $w_a$, however, belong to different models:
$q$ is sampled only in HoloEoS and $w_a$ only in CPL-VCDM.
They are not sampled jointly, so no $(q,w_a)$ posterior is shown.
The $\Lambda$CDM limits are indicated by reference markers in
the DE-parameter panels.

\begin{figure*}[!tp]
\centering
\includegraphics[width=0.98\textwidth]{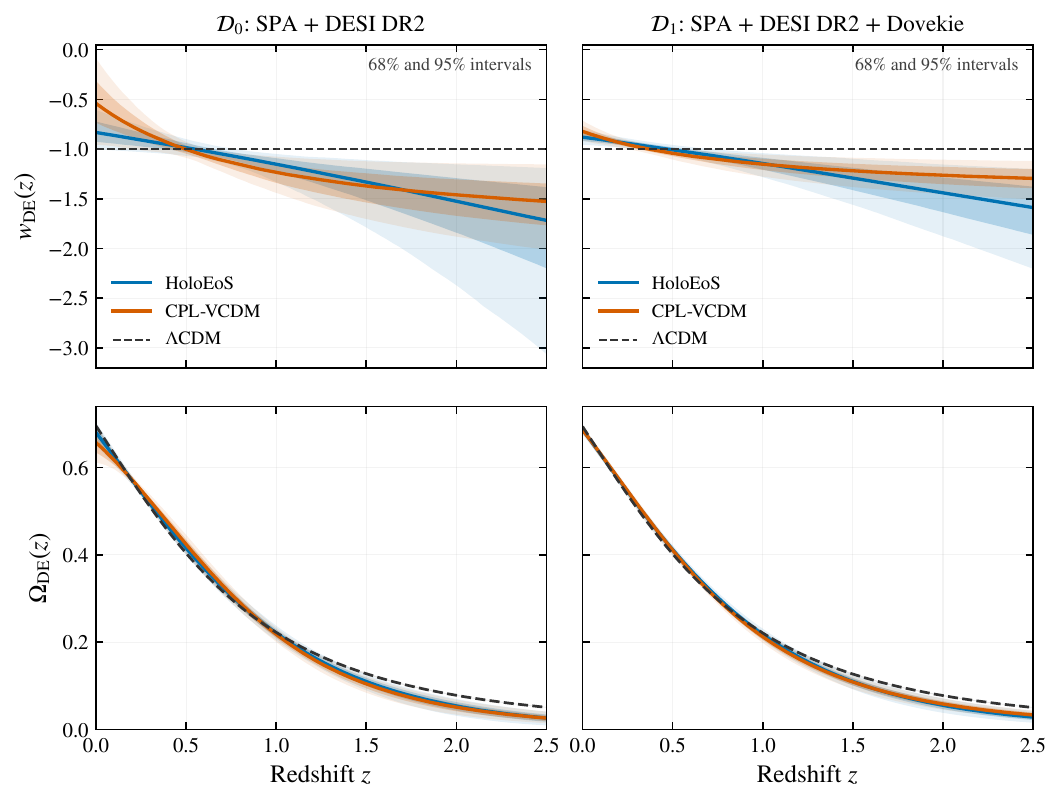}
\caption{Reconstructed DE histories for $\mathcal D_0$ (left) and
$\mathcal D_1$ (right).
Blue and orange correspond to HoloEoS and CPL-VCDM, respectively.
Solid curves show pointwise posterior medians, while dark and
light bands indicate central $68\%$ and $95\%$ credible intervals.
These intervals apply at each redshift and do not define
simultaneous bands for entire trajectories.
The top panels show $w_{\rm DE}(z)$, with the $\Lambda$CDM
value $-1$ indicated in gray. The bottom panels show 
$\Omega_{\rm DE}(z)$, with the thermal massive-neutrino 
contribution included in the total background energy density.
The gray curve and band show the $\Lambda$CDM posterior median
and central $68\%$ credible interval.
The phantom-divide crossing redshifts summarized in
Table~\ref{tab:crossing} are determined from individual
posterior samples.}
\label{fig:phantom_crossing_polychord}
\end{figure*}

Figure~\ref{fig:phantom_crossing_polychord} shows that the two
extensions yield similar DE density histories at low redshift
while allowing different equation-of-state evolution.
For CPL-VCDM with $w_a\ne0$, the phantom divide
$w_{\rm DE}=-1$ is crossed at
$a_\times=1+(1+w_0)/w_a$; a positive-redshift crossing requires
$0<a_\times<1$.
For HoloEoS, we determine the crossing redshift from each
reconstructed background trajectory.
Table~\ref{tab:crossing} summarizes the posterior crossing
probabilities and the crossing-redshift distributions conditional
on $0<z_\times<6$.
For the full data combination $\mathcal D_1$, the conditional
posterior medians and central $68\%$ credible intervals are
\begin{align}
 z_\times^{\rm HoloEoS}
 &=0.476^{+0.106}_{-0.081},\nonumber\\
 z_\times^{\rm CPL\text{-}VCDM}
 &=0.367^{+0.072}_{-0.059}.
 \label{eq:crossing_results}
\end{align}
These intervals overlap, although the median crossing redshifts differ.
The high crossing probabilities are conditional on the assumed
model and do not establish a preference for either extension over
$\Lambda$CDM. Interpreting them as a discovery significance 
would require a separately defined and calibrated statistical test.

% BEGIN EMBEDDED TABLE: crossing
\begin{table}[!tbp]
\centering
\caption{Posterior summaries of phantom-divide crossing.
For each model, $P_+$ and $P_6$ denote the posterior probabilities
of a crossing at $z_\times>0$ and $0<z_\times<6$, respectively.
The quoted medians and central $68\%$ credible intervals for
$z_\times$ are conditional on $0<z_\times<6$.
These probabilities describe crossing within each assumed model
and do not quantify a preference over $\Lambda$CDM.}
\label{tab:crossing}
\renewcommand{\arraystretch}{1.15}
\begin{ruledtabular}
\begin{tabular}{llccc}
Data & Model & $P_+$ & $P_6$ & $z_\times\mid(0,6)$ \\
\hline
$\mathcal D_0$ & HoloEoS & 0.9692 & 0.9692 & $0.554^{+0.105}_{-0.151}$ \\
$\mathcal D_0$ & CPL-VCDM & 0.9911 & 0.9907 & $0.495^{+0.064}_{-0.078}$ \\
$\mathcal D_1$ & HoloEoS & 0.9991 & 0.9986 & $0.476^{+0.106}_{-0.081}$ \\
$\mathcal D_1$ & CPL-VCDM & 0.9999 & 0.9999 & $0.367^{+0.072}_{-0.059}$ \\
\end{tabular}
\end{ruledtabular}
\end{table}
% END EMBEDDED TABLE: crossing

For the full data combination $\mathcal D_1$, HoloEoS and 
CPL-VCDM yield nearly identical clustering amplitudes,
$S_8\simeq0.835$ and $0.834$, respectively, compared with
$S_8\simeq0.822$ in $\Lambda$CDM.
Both extensions also favor slightly lower $H_0$ and higher
$\Omega_{m0}$ than $\Lambda$CDM.
For these data, neither extension therefore shifts $H_0$ upward
or $S_8$ downward relative to the reference model fitted to the
same data. These results illustrate that phantom-divide crossing alone
does not guarantee an alleviation of the $H_0$ or $S_8$ tension.
In particular, structure growth depends on both the background
expansion and the perturbation dynamics.

\subsection{Best-fit quality and model comparison}
\label{subsec:model_comparison}

The posterior constraints above describe the cosmological histories
allowed within each model. We now compare the best fits of HoloEoS
and CPL-VCDM with $\Lambda$CDM, accounting for the additional
parameters of the two extensions.

% BEGIN EMBEDDED TABLE: best_fits
\begin{table*}[!tbp]
\centering
\caption{Best physical-likelihood values reached by the completed
MH simulated-annealing searches, with all cosmological
and nuisance parameters varied and the final candidates freshly
re-evaluated, and the corresponding AIC differences. Differences are
relative to $\Lambda$CDM for the same data combination. Each extension
adds two parameters, so
$\Delta\mathrm{AIC}=\Delta\chi^2_{{\rm phys},{\rm best}}+4$.
The external Gaussian prior penalties are excluded. Differences are
computed from the full-precision values before rounding; displayed
digits identify the numerical candidates, not their optimization
accuracy. The stopping criterion is described in
Appendix~\ref{app:numerical_checks}.}
\label{tab:best_fits}
\begin{ruledtabular}
\begin{tabular}{llrrr}
Data & Model & $\chi^2_{{\rm phys},{\rm best}}$ & $\Delta\chi^2_{{\rm phys},{\rm best}}$ & $\Delta\mathrm{AIC}$ \\
\hline
$\mathcal D_0$ & $\Lambda$CDM & 615.343 & +0.000 & +0.00 \\
$\mathcal D_0$ & HoloEoS & 613.099 & -2.244 & +1.76 \\
$\mathcal D_0$ & CPL-VCDM & 612.190 & -3.153 & +0.85 \\
\hline
$\mathcal D_1$ & $\Lambda$CDM & 2259.501 & +0.000 & +0.00 \\
$\mathcal D_1$ & HoloEoS & 2252.103 & -7.398 & -3.40 \\
$\mathcal D_1$ & CPL-VCDM & 2251.404 & -8.097 & -4.10 \\
\end{tabular}
\end{ruledtabular}
\end{table*}
% END EMBEDDED TABLE: best_fits

Denoting the experimental likelihood by $\mathcal L_{\rm phys}$,
we define
\begin{equation}
 \chi^2_{\rm phys}=-2\ln\mathcal L_{\rm phys}.
 \label{eq:physical_chi2}
\end{equation}
This likelihood retains the normalization constants of all implemented
experimental terms, including the convention for analytic marginalization
over the supernova absolute-magnitude offset. The external Gaussian
information on $\tau$ and $T_{\rm cal}$ in
Eq.~(\ref{eq:external_gaussians}) enters the posterior analysis once,
through the normalized prior transform, but is not added to
$\mathcal L_{\rm phys}$. Accordingly, the likelihood best fits need not
coincide with posterior maxima or the marginalized medians reported above.

For each of the six model--data combinations, we perform a separate
MH simulated-annealing search of
$\chi^2_{\rm phys}$, varying all free cosmological and nuisance parameters.
The selected candidates are freshly re-evaluated with the final
implementation and the fixed experimental-likelihood normalization.
We denote the best values reached by $\chi^2_{{\rm phys},{\rm best}}$.
The annealing trajectories are used only for optimization; the
posterior constraints retain the PolyChord samples. The procedure
and its stopping criterion are summarized in
Appendix~\ref{app:numerical_checks}.

The Akaike information criterion (AIC)~\cite{Akaike:1974} combines
minus twice the maximized experimental log likelihood with the
penalty $2k$, where $k$ is the number of varied parameters.
Using the best values reached in the
annealing searches, we evaluate
\begin{align}
 \mathrm{AIC}
 &=\chi^2_{{\rm phys},{\rm best}}+2k,\nonumber\\
 \Delta\mathrm{AIC}
 &=\Delta\chi^2_{{\rm phys},{\rm best}}+2\Delta k
 =\Delta\chi^2_{{\rm phys},{\rm best}}+4,
 \label{eq:AIC_definition}
\end{align}
where $\Delta$ denotes the extension-minus-$\Lambda$CDM difference
for the same data combination. The explicitly varied cosmological
and nuisance parameters give $k=10$ for $\Lambda$CDM and $k=12$
for each extension. Counting the common analytically marginalized
supernova magnitude offset would add the same parameter penalty
to all $\mathcal D_1$ models and leave $\Delta\mathrm{AIC}$ unchanged.
The external Gaussian prior penalties do not enter the optimization
or this AIC calculation.

Table~\ref{tab:best_fits} lists the annealing best-fit statistics.
For $\mathcal D_0$, the physical-likelihood improvements relative to
$\Lambda$CDM are $\Delta\chi^2_{{\rm phys},{\rm best}}=-2.24$ for
HoloEoS and $-3.15$ for CPL-VCDM. The corresponding
$\Delta\mathrm{AIC}=+1.76$ and $+0.85$ show that these improvements
do not compensate for the penalty for two additional parameters.
With DESY5 Dovekie included in $\mathcal D_1$, the improvements become
$\Delta\chi^2_{{\rm phys},{\rm best}}=-7.40$ and $-8.10$, giving
$\Delta\mathrm{AIC}=-3.40$ and $-4.10$, respectively.
Thus the full data combination yields a lower AIC for both extensions
than for $\Lambda$CDM, with the two extensions separated by only
$0.70$ in AIC.

These modest AIC differences indicate comparable fitting performance
for HoloEoS and CPL-VCDM, without establishing a robust ordering
between the two extensions. They are neither posterior model
probabilities nor calibrated detection significances.

\section{Conclusions}
\label{sec:conclusions}

We have studied the two-parameter Hubble-running equation of state
$w_{\rm DE}=w_0(H/H_0)^q$ (HoloEoS) and realized its expansion
history within VCDM. The power-law dependence is motivated by holographic DE
constructions, but is not uniquely derived from a particular
infrared cutoff or entropy prescription.
For each regular expanding history, the potential of the
nondynamical auxiliary scalar can be reconstructed parametrically
from the background evolution and the standard-species density
and pressure. This specifies the gravitational perturbation
sector while retaining only the two tensor gravitational degrees
of freedom, with no independent propagating DE scalar.

For $q>0$ and $-1<w_0<0$, a regular trajectory connecting an early
radiation era to the present crosses the phantom divide at
$H/H_0=(-1/w_0)^{1/q}$.
Positive standard-species enthalpy fixes the direction of a simple
crossing from $w_{\rm DE}<-1$ to $w_{\rm DE}>-1$ as the Universe
expands. For positive DE density, this transition corresponds to
an inflection of the reconstructed potential, with
$V_{,\phi\phi}$ changing from positive to negative.
The VCDM constraint equations remain regular through the crossing;
the divergence of the effective adiabatic ratio
$P_{\rm DE}'/\rho_{\rm DE}'$ does not represent a physical scalar
propagation speed. On the radiation-dominated branch with
$w_0<0$ and $q>0$, both the DE density and the magnitude of its
pressure vanish toward the asymptotic past, despite
$w_{\rm DE}\to-\infty$, so their contributions relative to
radiation become negligible.

We constrained this realization using the combined SPT, Planck,
and ACT primary-CMB and lensing likelihoods with updated DESI
DR2 distances, and examined the effect of adding DESY5 supernovae
with Dovekie recalibration.
The comparison with $\Lambda$CDM and CPL-VCDM uses the same
standard-species and fixed massive-neutrino prescriptions.
For the full data combination $\mathcal D_1$, HoloEoS gives
$w_0=-0.881^{+0.040}_{-0.038}$ and
$q=0.453^{+0.143}_{-0.133}$, where the estimates are posterior
medians with central $68\%$ credible intervals.
The crossing redshifts, conditional on $0<z_\times<6$, are
$z_\times=0.476^{+0.106}_{-0.081}$ for HoloEoS and
$0.367^{+0.072}_{-0.059}$ for CPL-VCDM.
Both models favor histories that are phantom-like in the past
and quintessence-like today. Their high posterior crossing
probabilities describe evolution within each assumed model;
they are not discovery significances or model probabilities
relative to $\Lambda$CDM.

Adding the DESY5 supernova data with the Dovekie recalibration
tightens the cosmological constraints and brings the
HoloEoS and CPL-VCDM constraints on $H_0$, $\Omega_{m0}$,
and $S_8$ into closer agreement.
For $\mathcal D_1$, their posterior constraints favor slightly lower
$H_0$ and higher $S_8$ than the matched $\Lambda$CDM posterior.
Thus phantom-divide crossing alone does not guarantee an
alleviation of the $H_0$ or $S_8$ tension; the inferred clustering
also depends on the specified perturbation dynamics.

The MH simulated-annealing searches vary all cosmological and
nuisance parameters and minimize the physical $\chi^2$ without
adding external Gaussian prior penalties. For $\mathcal D_1$, the
best values reached improve relative to $\Lambda$CDM by
$\Delta\chi^2_{{\rm phys},{\rm best}}=-7.40$ for HoloEoS and $-8.10$
for CPL-VCDM. Including the penalty for two additional DE parameters
gives $\Delta\mathrm{AIC}=-3.40$ and $-4.10$, respectively.
The full data combination therefore gives both extensions a lower
AIC than $\Lambda$CDM, with comparable performance between them.
For $\mathcal D_0$ alone, the corresponding differences are
$+1.76$ and $+0.85$, so the smaller fit improvements do not overcome
the parameter penalty.

The central result is that a Hubble-scale-motivated DE history can
be realized within a specified gravitational theory that determines
both the background and cosmological perturbations, while achieving
fitting performance comparable to CPL-VCDM and allowing regular
phantom-divide crossing.
The quantitative constraints refer to the adopted priors, fixed
neutrino prescription, and numerical modeling choices.
Further constraints from structure growth, together with improved
nonlinear modeling in VCDM, can test whether the Hubble-running
history can be distinguished observationally from CPL-VCDM.

\appendix
\renewcommand{\theequation}{\Alph{section}.\arabic{equation}}

\section{Representative holographic backgrounds motivating HoloEoS}
\label{app:representativeHDE}

This appendix summarizes the defining density laws and equations
of state of representative holographic and agegraphic DE models
\cite{Wang:2016och}, supplementing the discussion in
Secs.~\ref{sec:introduction} and \ref{subsec:holoeos_motivation}.
We assume a spatially flat FLRW background in GR, with separately
conserved DE, pressureless matter, and radiation.
A prime denotes differentiation with respect to $\N=\ln a$.
These constructions provide background motivation for HoloEoS,
whose VCDM realization is tested in the observational analysis
of Sec.~\ref{sec:constraints}.
Table~\ref{tab:model_classification} compares their defining
prescriptions with HoloEoS.

\begin{table*}[!tbp]
\centering
\caption{Representative DE background prescriptions compared with HoloEoS.
The defining relations characterize the homogeneous evolution;
the propagating degrees of freedom depend on the underlying theory.
For HoloEoS, the perturbation dynamics and gravitational degrees
of freedom are fixed by its VCDM realization.}
\label{tab:model_classification}
\begingroup
\small
\renewcommand{\arraystretch}{1.12}
\setlength{\tabcolsep}{4pt}
\begin{ruledtabular}
\begin{tabular}{@{}p{0.20\textwidth}p{0.25\textwidth}p{0.18\textwidth}p{0.29\textwidth}@{}}
\RaggedCell{Construction} & \RaggedCell{Defining relation} & \RaggedCell{Background dependence} & \RaggedCell{Qualification} \\
\hline
\RaggedCell{Hubble-cutoff HDE \cite{Hsu:2004ri}} & \RaggedCell{$\rho_{\rm hde}\propto H^2$} & \RaggedCell{Local} & \RaggedCell{Tracks matter or radiation in the noninteracting setup} \\
\RaggedCell{Future-horizon HDE \cite{Li:2004rb}} & \RaggedCell{$\rho_{\rm hde}\propto L_f^{-2}$} & \RaggedCell{Future-integrated} & \RaggedCell{Requires the future-horizon boundary condition} \\
\RaggedCell{Particle-horizon HDE} & \RaggedCell{$\rho_{\rm hde}\propto L_p^{-2}$} & \RaggedCell{Past-integrated} & \RaggedCell{Does not accelerate in the noninteracting setup} \\
\RaggedCell{Ricci/Granda--Oliveros HDE \cite{Gao:2007ep,Granda:2008dk}} & \RaggedCell{$\rho_{\rm hde}\propto\alpha H^2+\beta\dot H$} & \RaggedCell{Local} & \RaggedCell{Depends on the derivative of the expansion rate} \\
\RaggedCell{Agegraphic/new agegraphic DE \cite{Cai:2007us,Wei:2007ty}} & \RaggedCell{$\rho_{\rm hde}\propto T^{-2}$ or $\eta^{-2}$} & \RaggedCell{Past-integrated} & \RaggedCell{Requires a choice of cosmological clock and its origin} \\
\RaggedCell{Tsallis/Barrow HDE \cite{Tavayef:2018xwx,Saridakis:2020zol}} & \RaggedCell{$\rho_{\rm hde}\propto H^p$ for $L=H^{-1}$} & \RaggedCell{Local} & \RaggedCell{Depends on the entropy and cutoff prescriptions} \\
\RaggedCell{Power-law future horizon} & \RaggedCell{$\rho_{\rm hde}\propto L_f^{-p}$} & \RaggedCell{Future-integrated} & \RaggedCell{Requires a normalization and a future boundary condition} \\
\RaggedCell{HoloEoS in VCDM} & \RaggedCell{$w_{\rm DE}=w_0(H/H_0)^q$} & \RaggedCell{Local background law} & \RaggedCell{Phenomenological; realized by reconstructing $V(\phi)$} \\
\end{tabular}
\end{ruledtabular}
\endgroup
\end{table*}
\subsection{Standard horizon cutoffs}

For the Hubble-radius cutoff $L=H^{-1}$, the density in Eq.~\eqref{eq:HDE_density} becomes
\begin{equation}
 \rho_{\rm hde}=3\chde^2\mpl^2H^2,\qquad
 \Ode=\chde^2.
\end{equation}
For $0<\chde^2<1$, separate conservation and
Eq.~\eqref{eq:hprimegeneral} give
$w_{\rm DE}=\Or/[3(1-\chde^2)]$.
Thus the component tracks radiation or pressureless matter rather than
driving acceleration \cite{Hsu:2004ri}. This limitation motivates
modifying either the cutoff or the power of the expansion scale.

The future-event-horizon prescription \cite{Li:2004rb} instead uses
\begin{align}
 \rho_{\rm hde}&=3\chde^2\mpl^2L_f^{-2},\nonumber\\
 L_f&=a\int_t^{t_{\rm end}}\frac{\dd\widetilde t}{a(\widetilde t)},
 \qquad \frac{1}{HL_f}=\frac{\sqrt{\Ode}}{\chde}.
\end{align}
Here $t_{\rm end}$ denotes the future endpoint of the cosmological
evolution. It is infinite if the evolution extends to arbitrarily
late cosmic times, and finite otherwise.
The future-horizon prescription requires the integral to converge. 
Since $\dot L_f=HL_f-1$, conservation yields
\begin{equation}
 w_{\rm DE}=-\frac{1}{3}-\frac{2}{3\chde}\sqrt{\Ode}.
\end{equation}
Accelerating solutions are possible, but the local evolution equation
must be supplemented by the boundary condition selecting the future
horizon \cite{Li:2012xf,Lin:2021hde}.
For the particle horizon $L_p=a\int_0^t\dd\widetilde t/a(\widetilde t)$,
where the lower limit is the initial cosmological boundary,
$\dot L_p=HL_p+1$ instead gives
\begin{equation}
 w_{\rm DE}=-\frac{1}{3}+\frac{2}{3\chde}\sqrt{\Ode}.
\end{equation}
This component cannot produce acceleration when combined with
noninteracting dust and radiation.

\subsection{Local curvature and agegraphic prescriptions}

Ricci HDE \cite{Gao:2007ep} and the Granda--Oliveros model
\cite{Granda:2008dk,Granda:2008tm} prescribe
\begin{equation}
 \rho_{\rm hde}=3\mpl^2(\alpha H^2+\beta\dot H),
 \qquad \Ode=\alpha+\beta\frac{H'}{H},
\end{equation}
with dimensionless constants $\alpha$ and $\beta$. The Ricci prescription $\rho_{\rm hde}\propto R$, with
$R=6(2H^2+\dot H)$, corresponds to $\alpha=2\beta$.
For $\beta\ne0$ and $\Ode>0$, Eq.~\eqref{eq:hprimegeneral} implies
\begin{equation}
 w_{\rm DE}=-\frac{1}{3\Ode}
 \left[3+\Or+\frac{2}{\beta}(\Ode-\alpha)\right].
\end{equation}
This is an example of a local expansion-dependent density that
induces a time-dependent equation of state.

The original and new agegraphic DE models prescribe the DE density
in terms of the cosmic age $T$ and the conformal age $\eta$,
respectively \cite{Cai:2007us,Wei:2007ty}. For a dimensionless
constant $n>0$, these densities are
\begin{align}
 \rho_{\rm hde}&=3n^2\mpl^2T^{-2},
 &T&=\int_0^t\dd\widetilde t,\\
 \rho_{\rm hde}&=3n^2\mpl^2\eta^{-2},
 &\eta&=\int_0^t\frac{\dd\widetilde t}{a(\widetilde t)}.
\end{align}
Their equations of state are, respectively,
\begin{align}
 w_{\rm DE}&=-1+\frac{2}{3n}\sqrt{\Ode},\\
 w_{\rm DE}&=-1+\frac{2e^{-\N}}{3n}\sqrt{\Ode}.
\end{align}
These prescriptions replace the horizon length by a cosmological
clock; neither directly yields Eq.~\eqref{eq:HubbleRunningEoS}.

\subsection{Entropy-deformed power laws}

For the Hubble-radius infrared cutoff $L=H^{-1}$, the Tsallis
and Barrow HDE densities take the respective forms
\cite{Tavayef:2018xwx,Barrow:2020tzx,Saridakis:2020zol}
\begin{equation}
 \rho_{\rm hde}=B H^{4-2\delta},\qquad
 \rho_{\rm hde}=C H^{2-\Delta},
\end{equation}
The positive normalizations $B$ and $C$ generally carry
dimensions, and the constant dimensionless exponents $\delta$
and $\Delta$ characterize the entropy deformations. Their mass dimensions are $[B]=2\delta$ and
$[C]=2+\Delta$, so both densities have dimension four.
The standard $H^2$ dependence is recovered at
$\delta=1$ or $\Delta=0$, respectively.
More generally, for a constant exponent $p$, separate conservation
of a density $\rho_{\rm hde}\propto H^p$ gives
\begin{equation}
 w_{\rm DE}=-1-\frac{p}{3}\frac{H'}{H}.
\end{equation}
A related prescription uses a power of the future horizon:
\begin{equation}
 \rho_{\rm hde}=B_f L_f^{-p},\quad
 w_{\rm DE}=-1+\frac{p}{3}\left(1-\frac{1}{HL_f}\right).
\end{equation}
Here $B_f>0$ is a normalization constant distinct from $B$ above,
with mass dimension $[B_f]=4-p$.
For fixed $B_f$ and $p\ne0,2$, the ratio $1/(HL_f)$ depends
explicitly on both $H$ and $\Ode$.
At $p=2$, its explicit dependence on $H$ cancels.
The local evolution equation for $L_f$ must be supplemented
by the boundary condition selecting the future event horizon.
For $p=0$, the prescription reduces to a constant DE density,
$\rho_{\rm hde}=B_f$, independent of $L_f$.

These examples motivate a power-law dependence on the expansion scale,
but a power law for the density does not imply the HoloEoS power law
for $w_{\rm DE}$. Equation~\eqref{eq:HubbleRunningEoS} is a separate
phenomenological ansatz; its VCDM realization is specified in
Sec.~\ref{subsec:VCDMreconstruction}.

\section{VCDM linear cosmological perturbations and initial conditions}
\label{app:vcdm_perturbations}

This appendix summarizes the treatment of linear scalar perturbations
and their initial conditions in the VCDM realization used in
Secs.~\ref{subsec:holoeos_validity} and \ref{subsec:run_settings}.
Only in this appendix, a prime denotes $\dd/\dd\tau$, where $\tau$
is conformal time, unrelated to the reionization optical depth
denoted by the same symbol in the main text.
An overdot retains the cosmic-time meaning used
in the main text. Thus $\dd t=a\,\dd\tau$, $\mathcal H=aH$, and
$\dot H\equiv\dd H/\dd t=H'/a$.

In the unitary-gauge formulation, the constraint
$\partial_i\phi=0$ imposed by the Lagrange multiplier requires
the perturbation $\delta\phi$ of the auxiliary scalar field
to vanish for inhomogeneous Fourier modes (wavenumber $k>0$).
We use the remaining scalar spatial diffeomorphism to
set the potential $E$ in the spatial-metric contribution
$2\partial_i\partial_jE$ to zero. This metric potential is
distinct from the normalized Hubble rate $E$ in the main text.
The scalar metric then reads
\begin{align}
 \dd s^2=a^2\Bigl[&-(1+2A)\dd\tau^2
 +2\partial_iB\,\dd\tau\dd x^i\nonumber\\
 &+(1-2\psi)\delta_{ij}\dd x^i\dd x^j\Bigr]\,,
 \label{eq:vcdm_scalar_metric}
\end{align}
where $A$, $B$, and $\psi$ are metric perturbations.
The scalar shift $B$ generally remains: there is no independent
local temporal gauge freedom with which to impose $B=0$ as well.
We introduce the combinations
\begin{equation}
 \Psi=A+\mathcal H B+B',\qquad
 \Phi=\psi-\mathcal H B.
 \label{eq:vcdm_bardeen_variables}
\end{equation}
These are the usual GR Bardeen combinations in the spatial gauge
$E=0$. Their use is a change of perturbation variables, not the
imposition of a diagonal metric in the preferred foliation.
The shift and additional auxiliary perturbations are determined
by the constraints; they do not introduce an independent
propagating gravitational scalar
\cite{VCDMFoundation2020,VCDMArora2025}.
The matter perturbations below are expressed in the corresponding
longitudinal combinations as well.

We use Fourier modes $e^{i\bm k\cdot\bm x}$, so
$\partial_j\to ik_j$ and $k^2=\delta^{ij}k_i k_j$.
For each standard species $I$, 
we define its longitudinal velocity potential $V_I$ through
$(\delta T^0{}_{j})_{I,\mathrm L}=(\rho_I+P_I)\partial_j V_I$.
The velocity divergence used below is
\begin{equation}
 \theta_I\equiv
 \frac{ik^j(\delta T^0{}_{j})_{I,\mathrm L}}{\rho_I+P_I}
 =-k^2V_I,
\end{equation}
with spatial indices raised by $\delta^{ij}$. This is the usual
velocity-divergence convention used in \texttt{CLASS}
\cite{MaBertschinger1995,Blas:2011rf}.
Here $\mathrm L$ labels the longitudinal combination, rather than
an additional gauge choice in VCDM. In the $E=0$ variables of
Eq.~\eqref{eq:vcdm_scalar_metric}, defining the momentum potential by
$(\delta T^0{}_{j})_I=(\rho_I+P_I)\partial_j v_I$ gives
$V_I=v_I-B$. These momentum definitions apply to collisionless
species as well as to perfect fluids.

We retain the physical background densities and pressures of the
main text, with $\rho_s+P_s=\sum_I(\rho_I+P_I)$.
In these variables, the exact VCDM momentum constraint is
\cite{VCDMArora2025}
\begin{equation}
 \Phi'+\mathcal H\Psi=
 \frac{\left(k^2-3a^2\dot H\right)
 \sum_I(\rho_I+P_I)\theta_I}
 {k^2\left[2\mpl^2k^2/a^2+3(\rho_s+P_s)\right]}.
 \label{eq:phi_evolution_vcdm}
\end{equation}
The sum includes all standard species and excludes the effective DE
sector. No density or pressure is rescaled by $3\mpl^2$ in this form.
In particular, massive neutrinos are retained through their thermal
density and pressure, as in Sec.~\ref{subsec:run_settings}.
The remaining metric and standard-species equations have their usual
GR form, including the standard-species anisotropic stress. 
No independent perturbation or sound
speed is assigned to an effective DE fluid.

Using the background identity
$\dot H=-[\rho_s+P_s+\rho_{\rm DE}+P_{\rm DE}]/(2\mpl^2)$,
Eq.~\eqref{eq:phi_evolution_vcdm} can equivalently be written as
\begin{align}
 \Phi'+\mathcal H\Psi
 &=\frac{a^2}{2\mpl^2k^2}
 \sum_I(\rho_I+P_I)\theta_I\,\mathcal C(k,\tau),
 \label{eq:momentum_GR_limit}\\
 \mathcal C(k,\tau)
 &=1+\frac{3(\rho_{\rm DE}+P_{\rm DE})}
 {2\mpl^2k^2/a^2+3(\rho_s+P_s)}.
 \label{eq:vcdm_momentum_correction}
\end{align}
Here $\rho_s$ and $P_s$ are the total standard-species density and
pressure defined in Sec.~\ref{subsec:VCDMreconstruction}.
The departure from the GR momentum constraint is encoded in
$\mathcal C-1$.
For $k>0$ and $\rho_s+P_s>0$, the denominator
$2\mpl^2k^2/a^2+3(\rho_s+P_s)$ in
Eq.~\eqref{eq:vcdm_momentum_correction} is positive.
Thus $\mathcal C$ remains finite at a regular phantom-divide
crossing and equals unity at $w_{\rm DE}=-1$.
It is identically unity in $\Lambda$CDM and approaches unity
as $k/a\to\infty$ for a fixed regular background.

In VCDM, standard-matter perturbations remain dynamical even though
there is no propagating gravitational scalar.
In the high-wave-number limit, their no-ghost conditions and
propagation speeds coincide with those of the same minimally
coupled matter in GR. The two tensor polarizations of 
gravitational waves propagate
at the speed of light, $c_T=1$ \cite{VCDMFoundation2020}.
The effective DE adiabatic ratio in
Eq.~\eqref{eq:adiabaticSoundSpeed} does not represent a physical
DE propagation speed. Its divergence at phantom-divide crossing
therefore does not by itself signal an instability.

For $w_0<0$ and $q>0$, the HoloEoS density and pressure are suppressed
in the asymptotic radiation era according to
Eq.~\eqref{eq:rho_superexponential}. At the prior boundaries $q=0$
or $w_0=0$, their ratios to radiation still vanish, although their
absolute values need not vanish. CPL-VCDM has the same negligible
early-time DE fraction under the prior condition $w_0+w_a<0$.
In each case,
$(\rho_{\rm DE}+P_{\rm DE})/(\rho_s+P_s)\to0$, so
Eq.~\eqref{eq:vcdm_momentum_correction} approaches its GR limit
also for superhorizon modes. We therefore use the standard regular
adiabatic initial conditions implemented in \texttt{CLASS} for
photons, baryons, CDM, and neutrinos. There is no independent DE
scalar initial condition; auxiliary perturbations follow from the
constraints.

\section{Likelihood conventions and numerical optimization}
\label{app:numerical_checks}

The experimental likelihood $\mathcal L_{\rm phys}$ contains the
CMB, updated DESI DR2, and, for $\mathcal D_1$, supernova contributions,
with the normalization constants of the implemented likelihoods.
The external Gaussian information on $\tau$ and $T_{\rm cal}$ enters
posterior inference once through the normalized prior transforms.
The associated penalties are excluded from the physical-likelihood
minimization and the AIC. The same nuisance-parameter counting and
analytically marginalized supernova magnitude convention are used
for all models within a given data combination.
With the adopted fixed supernova covariance, analytic marginalization
and profiling over the common magnitude offset differ only by a
model-independent normalization, which cancels from the reported
likelihood and AIC differences.

The final CLASS v2 implementation uses cubic Hermite interpolation
of the logarithmic HoloEoS DE density and common numerical domain
checks. Comparisons with reconstructed earlier implementations
covered 42 archived parameter vectors. The $\Lambda$CDM and
CPL-VCDM likelihoods agreed at the tested points; at the selected
maximum-weight HoloEoS posterior points, the absolute changes in
$\ln\mathcal L_{\rm phys}$ were $2.87\times10^{-5}$ for $\mathcal D_0$
and $8.31\times10^{-4}$ for $\mathcal D_1$. All tested admission
decisions agreed. These pointwise checks support retaining the
original weighted posterior samples; they do not constitute an
independent reconstruction of the posterior distributions.

For the AIC comparison, dedicated MH simulated-annealing searches
were performed for all six model--data combinations with the fixed
final CLASS v2 implementation. The procedure follows the MontePython
annealing implementation recorded with the optimization
outputs.\footnote{\href{https://github.com/GuillermoFrancoAbellan/MontePython/blob/main/run_minimizer.sh}{MontePython \texttt{run\_minimizer.sh} implementation.}}
All cosmological and nuisance parameters were varied, excluding the
external Gaussian penalties on $\tau$ and $T_{\rm cal}$ from the
objective function. The final candidates were freshly re-evaluated
with the common likelihood convention. Several initial seeds were
checked, with an adopted operational stability threshold of $0.3$
in $\chi^2_{\rm phys}$. This stopping criterion does not certify a
global minimum or a statistical uncertainty on the reported AIC
differences. The digits in Table~\ref{tab:best_fits} identify the saved
candidates, with differences calculated before rounding.

These likelihood-optimized points need not coincide with posterior
modes or marginalized medians, because the external Gaussian
information enters only the posterior analysis. The annealing
trajectories do not enter the posterior constraints, figures, or
crossing probabilities.

\section*{Data Availability}
The public data products are described in the cited release papers
for Planck, ACT, SPT, DESI, and DES/Dovekie. The authors can provide
the modified CLASS code, weighted posterior samples, run
configurations, numerical tables, figure scripts, annealing best-fit
parameter vectors, physical-likelihood summaries, and numerical
validation records upon reasonable request.

\begin{acknowledgments}
A.D.F.\ gratefully acknowledges the hospitality and financial support provided by Mahidol University during his visit. This visit helped inspire the present research project.
S.T.\ acknowledges support from JSPS KAKENHI Grant Nos.~26K07090 and 26H00847, and from the Waseda University Special Research Projects (No.~2026C-486). S.T.\ also thanks Mahidol University for its kind hospitality during his visit.
The work of J.Z.\ was supported by the Japanese Government (MEXT) Scholarship Program. The authors acknowledge the use of {\tt Maple} (for computer algebra calculations) and ChatGPT-6 (OpenAI, version dated October 2026) to assist in drafting and proofreading sections of the manuscript and for exploratory algebraic/numerical checks; all scientific content, derivations, and conclusions were verified by the authors, who take full responsibility for the final text. The numerical calculations presented in this work were performed in part using the Heian computing system at the Yukawa Institute for Theoretical Physics, Kyoto University.
\end{acknowledgments}

% Bibliography format follows astroph_v1.tex: explicit bibitems and article titles.

\end{document}